\documentclass{aa}
\usepackage{longtable}
\usepackage{multirow}
\usepackage{booktabs,fixltx2e}
\usepackage[flushleft]{threeparttable}
\usepackage{graphicx,natbib,txfonts,color}
\bibpunct{(}{)}{;}{a}{}{,}

\begin{document}

\title{Chemical abundance study of two strongly \textit{s}-process enriched post-AGB stars in the LMC: J051213.81-693537.1 and J051848.86-700246.9
\thanks{Based on observations collected with the Very Large Telescope
  at the ESO Paranal Observatory (Chili) of programme number 088.D-0433.} }

\author{K. De Smedt\inst{1}
\and H. Van Winckel\inst{1}
\and D. Kamath\inst{1}
\and P. R. Wood\inst{2}
}

\offprints{K. De Smedt, kenneth.desmedt@ster.kuleuven.be}

\institute{ Instituut voor Sterrenkunde, K.U.Leuven, Celestijnenlaan 200D,
B-3001 Leuven, Belgium,
\and  Research School of Astronomy and Astrophysics, Mount Stromlo Observatory, 
Weston Creek ACT 2611, Australia
}

\date{Received  / Accepted}

\authorrunning{K. De Smedt et al.}
\titlerunning{Chemical abundance study of two LMC \textit{s}-process enriched post-AGB stars}

\abstract
{This paper is part of a larger project in which we systematically study the chemical abundances 
of extra-galactic post-asymptotic giant branch (post-AGB) stars. The
aim of our programme is to derive chemical abundances of stars
covering a large range in luminosity and metallicity with the ultimate goal of
testing, constraining and improving our knowledge of the poorly understood AGB
phase,  especially the third dredge-up mixing processes and associated \textit{s}-process nucleosynthesis.}
{Post-AGB photospheres are dominated by atomic lines and indicate the effects of internal 
chemical enrichment processes over the entire stellar lifetime. 
In this paper, we study two carefully selected post-AGB stars: J051213.81-693537.1 and J051848.86-700246.9
in the Large Magellanic Cloud (LMC). Both objects show signs of \textit{s}-process enhancement. The combination of 
favourable atmospheric parameters for detailed abundance studies and
their known distances (and hence luminosities and initial masses) 
make these objects ideal probes of the AGB third dredge-up and \textit{s}-process
nucleosynthesis in that they provide observational constraints for theoretical
AGB models.}
{We use high-resolution optical UVES spectra to determine accurate
  stellar parameters and subsequently perform detailed elemental
  abundance studies of post-AGB stars. 
Additionally, we use available photometric data covering optical and IR bands to construct spectral energy
 distributions for reddening and luminosity determinations. We then estimate initial masses from
 theoretical post-AGB tracks.}
{ We obtained accurate atmospheric parameters for
  J051213.81-693537.1 (T$_{\textrm{eff}}$ = 5875 $\pm$ 125 K, $\log\,g$ = 1.00 $\pm$
  0.25 dex, [Fe/H]= $-$0.56 $\pm$ 0.16 dex) and
  J051848.86-700246.9 (T$_{\textrm{eff}}$ = 6000 $\pm$ 125 K, $\log\,g$ = 0.50 $\pm$
  0.25 dex, [Fe/H]= $-$1.06 $\pm$ 0.17 dex). Both stars show extreme \textit{s}-process enrichment associated
 with relatively low C/O ratios of 1.26 $\pm$ 0.40 and 1.29 $\pm$ 0.30  for J051213-693537.1 and J051848-700246.9, respectively.  
 We could only derive upper limits of the lead (Pb)
 abundance. These upper limits show a possible very slight Pb overabundance with respect 
 to heavy \textit{s}-elements for J051213-693537.1, while J051848-700246.9 shows an upper limit
 of the Pb abundance similar to [hs/Fe]. A comparison with
 theoretical post-AGB evolutionary tracks in the HR-diagram reveals that both stars have low initial
 masses between 1.0 and 1.5 M$_{\odot}$. 
}
{This study adds to the results obtained so far on
a very limited number of \textit{s}-process enriched stars in the
Magellanic Clouds. 
With the addition of the two stars in this study, we find an increasing 
discrepancy between observed and predicted Pb abundances towards lower metallicities 
of the studied \textit{s}-process rich post-AGB stars in the Magellanic Clouds. 
The more metal-rich J051213-693537.1 fits 
 the theoretical Pb abundance predictions well, while the five other objects 
with [Fe/H] < 1, including J051848-700246.9, have much lower Pb overabundances 
than predicted. In all objects found so far, including the objects in this
study, the C/O ratio is very moderate because of the enhancement of O as well as C. 
We find that all \textit{s}-process rich stars in the LMC and SMC
studied so far, cluster in the same region of the HR-diagram and
are associated with low-mass stars with a low
metallicity on average. We corroborate the published lack of
correlation between the metallicity and the neutron irradiation, while
the neutron exposure ([hs/ls]) is strongly correlated with the third
dredge-up efficiency ([s/Fe]). These correlations seem to hold in
our Galaxy as well as in the Magellanic Clouds.
}


\maketitle


\section{Introduction}\label{sect:intro}

Post-asymptotic giant branch (post-AGB) stars are low- to 
intermediate-mass stars (M\,$\leqslant$\,8M$_{\odot}$) in a transient evolutionary phase 
between the asymptotic giant branch (AGB) and the white dwarf (WD) phase. 
In single stars, a superwind mass loss terminates the AGB phase after which the star 
evolves onto the post-AGB phase, eventually cooling down as a white dwarf. 
During the AGB phase, the end products of internal chemical processes 
like C,N,O and \textit{s}-process elements are 
transported to the stellar surface by multiple dredge-up events, called the third 
dredge-up (TDU). The \textit{s}-process synthesis in AGB stars is an important contributor to the 
cosmic abundances past the iron peak and these stars are also expected to 
be important contributors to the cosmic carbon and nitrogen enrichment 
\citep[e.g][]{romano10,kobayashi11}.

Photospheric spectra of AGB stars are dominated by molecular bands
and they are often veiled by circumstellar dust, making it difficult to study abundances of individual elements 
\citep[e.g.][]{abia08}. However, post-AGB spectra are dominated by atomic transitions, allowing
extensive abundance studies of individual elements. Their photospheres bear witness to the 
total chemical changes accumulated
during the stellar lifetime, making them ideal tracers 
for the study of AGB nucleosynthesis and associated TDUs. 

Chemical studies of Galactic post-AGB stars show a large variety
of chemical abundance patterns \citep[e.g.][and references
therein]{vanwinckel03}. Only about 25\% of the studied Galactic 
post-AGB stars indicate strong \textit{s}-process enrichment
\citep[e.g.][and references therein]{rao11}. Although stars in the Galactic sample of post-AGB stars are well studied, 
their unknown distances hamper accurate luminosity determinations and hence 
accurate initial mass estimates, which are a key parameter in constraining single star evolution 
and nucleosynthesis AGB models. 

One way to overcome this distance problem, is to study post-AGB stars in the Magellanic Clouds.
The distances to the Magellanic Clouds are well known, with the LMC at
a distance of about 50 kpc \citep{keller06,reid10,storm11} and the SMC at a distance
of about 60 kpc \citep{keller06}. Furthermore, the Magellanic 
Cloud stars generally have lower mean metallicities than their
Galactic counterparts, with a mean metallicity of $\sim$\,-0.5\,dex for the
LMC \citep[e.g.][]{geisler09,lapenna12} and $\sim$\,-0.7\,dex for the
SMC \citep{luck98}. This allows for 
the study of post-AGB stars in a broad metallicity range. 

Therefore, we initiated a larger project in which we search for post-AGB
stars in the Magellanic Clouds and then use these objects to study the
poorly understood AGB TDUs and \textit{s}-process nucleosynthesis.
\citet{vanaarle11} and \citet[][accepted]{kamath15_accepted}, provide catalogues of 
spectroscopically verified optically visible post-AGB candidates in the Large Magellanic
Cloud (LMC) and \citet{kamath14} provides a catalogue of
spectroscopically verified optically visible post-AGB candidates in the Small Magellanic Cloud (SMC).

In this paper, we present the detailed chemical study of two newly
identified LMC post-AGB candidates that were
carefully selected (based on their stellar parameters and spectral
features) from our optically visible post-AGB LMC catalogues.

Our previous abundance studies of strongly \textit{s}-process enriched post-AGB
stars in the Magellanic Clouds by \citet{desmedt12},
\citet{vanaarle13}, and \citet{desmedt14} show that there are still
large discrepancies between observed abundances and theoretical
predictions. The biggest discrepancies are related to the C/O ratio since
observationally derived C/O ratios are small in comparison to the theoretical
predictions. The theoretical carbon abundance is often overestimated,
whereas the derived oxygen abundance is larger than predicted.  Also
lead (Pb, Z=82), which is considered to be the end product of the
\textit{s}-process chain due to its double magic number, is strongly
overestimated by theoretical models as shown in \citet{desmedt14}.

We describe the observations and data reduction of both objects in Sect. \ref{sect:obs}. 
Detailed abundance analysis of the obtained spectra is presented in Sect. \ref{sect:spec} followed by 
the discussion of the abundances in Sect. \ref{sect:abun}. Luminosity determinations and initial mass 
estimates are presented in \ref{sect:sed}. We discuss the neutron irradiation of our sample stars in Sect. \ref{sect:neutron}.
Finally, we end with a brief discussion and conclusions in Sect. \ref{sect:conc}.

\section{Observations and data reduction}\label{sect:obs}

We obtained high-resolution spectra using the UVES echelle spectrograph
\citep{dekker00} mounted on the 8m UT2 Kueyen
Telescope of the VLT array at the Paranal Observatory of ESO in
Chili. We selected the dichroic beam-splitter resulting in 
a wavelength coverage for the blue arm 
from approximately 3280 to 4530 \AA{}, and in the red arm for the 
lower and upper part of
the mosaic CCD chip from approximately 4780 to 5770 \AA{} and
from 5800 to 6810 \AA,{} respectively. A slit width of 1 arcsecond 
was used as a compromise between spectral resolution 
and slit-loss minimalisation. An overview of the observations is given in
Table \ref{table:obs}. For J051848.86-700246.8 (hereafter abbreviated to
J051848), only one blue spectrum was available because of an instrumental error.
For J051213.81-693537.1 (hereafter abbreviated to
J051213), two additional spectra were also obtained but because of the 
significant lower quality with respect to the six other spectra, these 
two spectra are not used for this study. The spectra of each object were 
obtained in one night and in one large time interval. 

\begin{table}[tb!]
\caption{\label{table:obs} Overview of observations. 'Exp. Time represents the exposure time. 
The red arm is the combination of the lower and upper part of
the mosaic CCD chip. The parameter $V$ represents the visual magnitude.}
\begin{center}
\begin{tabular}{lcccc} \hline\hline
Object name & Exp. time (s) & Exp. time (s)  &  $V$  \\
(IRAC)      & Blue arm    & Red arm          &  mag \\
\hline
J051213.81-693537.1 & 6 $\times$ 1500 & 6 $\times$ 1500   &  15.7  \\
J051848.86-700246.8 & 1 $\times$ 1500 & 2 $\times$ 1500   &  15.5  \\
\hline
\end{tabular}
\end{center}
\end{table}

The data was reduced using the UVES pipeline (version 5.3.0) in the Reflex environment of ESO \footnote{https://www.eso.org/sci/software/reflex/}.
This reduction includes the standard steps of extracting frames,
determining wavelength calibration, applying this scale to 
flat-field divided data and cosmic-clipping. The standard reduction parameters of the UVES pipeline 
were used as these gave the best signal-to-noise (S/N) of the final spectra.

After reduction, weighted mean spectra were calculated for the three
wavelength ranges. Since all spectra were taken in small time
intervals, no velocity corrections were needed for calculating the
weighted mean. These weighted mean spectra were then divided into
subspectra, each with a fixed wavelength range of 120 \AA{} for which the
first and last 10 \AA{} overlapped with the previous and subsequent
spectra.  Each of these subspectra was normalised individually by
fitting fifth order polynomials through interactively defined
continuum points.  All subspectra were then again merged into one
large spectrum, which is used for the spectral analysis. In the
wavelength regions where the subspectra overlapped, the mean flux was
calculated.

The wavelength range from 3280
to about 3900 \AA{} has a poor S/N, making it unsuitable for accurate spectral analysis. 
We therefore did not use this
spectral region in the analysis. The mean S/N in the remaining
wavelength range of the blue spectrum is about 25 for both stars. The
red spectra have a higher mean S/N between 80 and 100. The high
resolution of the UVES spectra permits accurate radial velocity
determinations  using the positions of spectral lines. Line
positions are measured by fitting a Gaussian through the measured line
profiles, line identification is based upon the rest wavelengths from
the VALD database \citet{vald}. We find a radial velocity of 227.8
$\pm$ 1.9 km/s for J051848 and a radial velocity of 284.3 $\pm$ 1.0
km/s for J051213. Both radial velocities are consistent with the
average radial velocity of the LMC, which is approximately 270 km/s
\citep{vandermarel02}.

\section{Spectral analyses}\label{sect:spec}

\begin{figure}[t!]
\resizebox{\hsize}{!}{\includegraphics{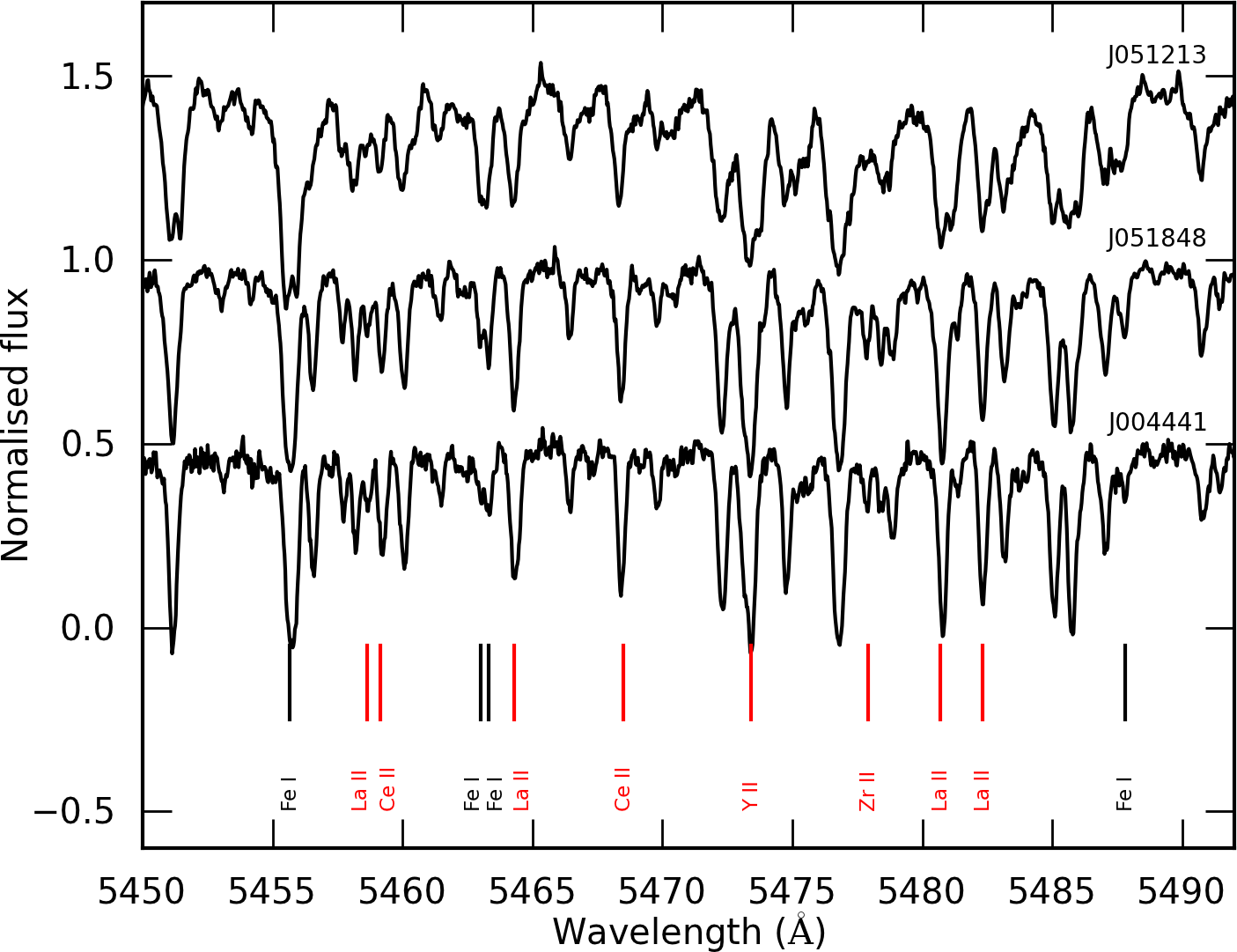}}
\caption{Comparison of the normalised spectra of J051213 (upper), J051848 (middle), and J004441 (lower). 
The upper and lower spectra have been shifted in flux for clarity and all spectra are shifted to a zero velocity. 
Red and black vertical lines mark positions of \textit{s}-nuclei and non \textit{s}-nuclei, respectively. See text for more information.}\label{fig:comp1}
\end{figure}
\begin{figure}[t!]
\resizebox{\hsize}{!}{\includegraphics{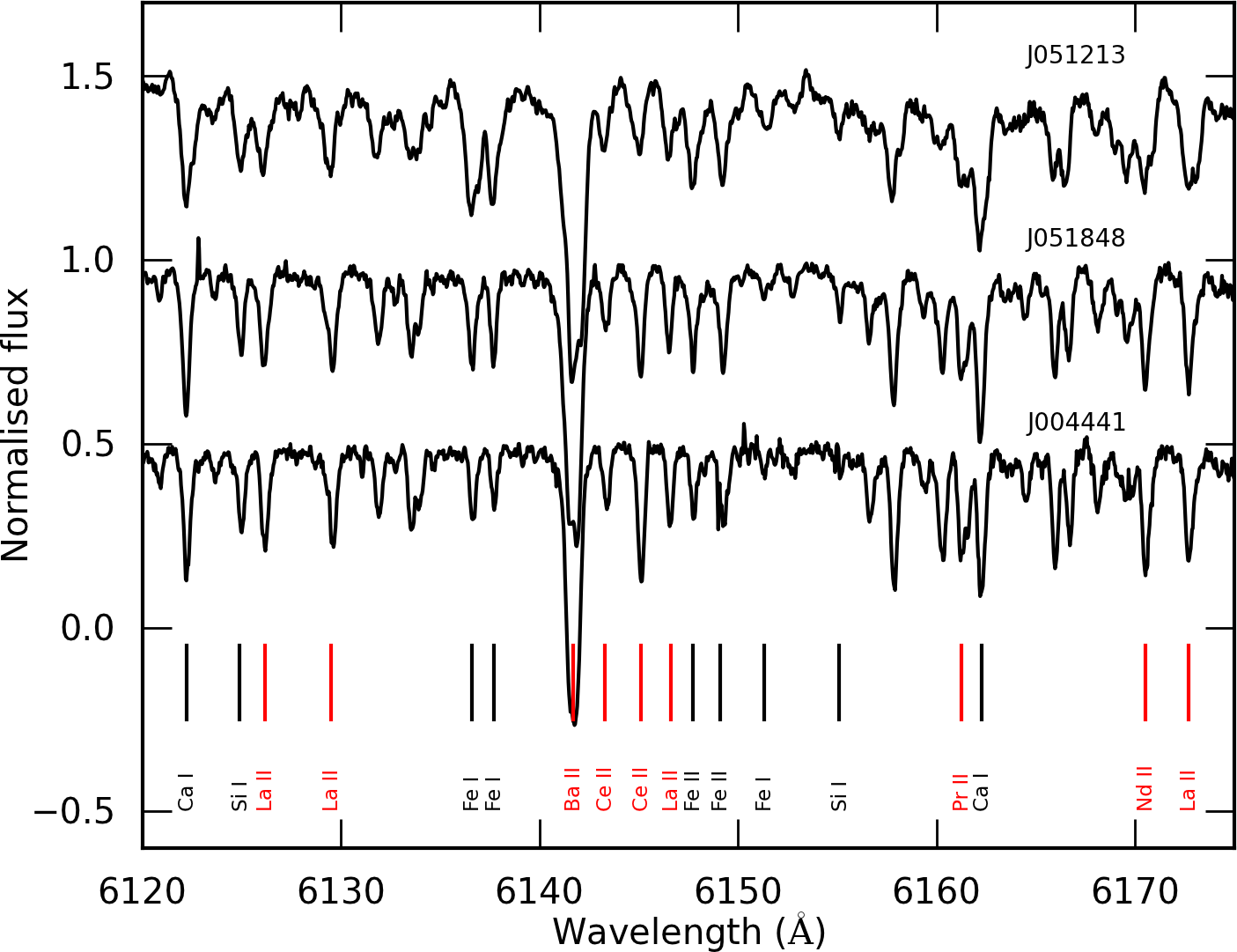}}
\caption{Similar to Fig. \ref{fig:comp1}, except for another wavelength range.}\label{fig:comp2}
\end{figure}

A visual inspection of the UVES spectra immediately shows the very
strong \textit{s}-process enrichment for both objects. 
Figs. \ref{fig:comp1} and \ref{fig:comp2} show different wavelength
regions of J051213 and J051848. Both figures also show the 
corresponding spectral region of J004441.04-732136.4 (J004441), which is a 
strongly \textit{s}-process enhanced SMC post-AGB star \citep{desmedt12}.
All three objects have similar atmospheric parameters except for the metallicity as described in Sect. \ref{subsect:atmos}. Both J051213 and J051848 show strong 
Ba II (Fig \ref{fig:comp2}) and Y II  (Fig \ref{fig:comp1}) lines, which is a
clear indication of \textit{s}-process enrichment. Although J051213 and J051848 are members of the LMC, while J004441 is an SMC object, they have similar spectra. 


The spectral analyses includes atmospheric parameter and
abundance determination.  We use PyMOOG, our own Python wrapper 
around the local thermal equilibrium (LTE)
abundance calculation routine MOOG \citep[version June 2014][]{sneden73}. 
For the analyses
of our two LMC stars, we use the LTE Kurucz-Castelli atmosphere models
\citep{castelli04}.  As with the radial-velocity determination,
spectral lines are identified using linelists from the VALD database
\citep{vald}.  We combine the VALD line-lists with a list of lines  gathered at the Instituut voor Sterrenkunde for the chemical
analysis of A, F, and G stars \citep{vanwinckel00}. The line-lists implemented in PyMOOG
cover a wavelength range from 3000 up 11000 \AA{}. This covers the
full wavelength coverage of the UVES spectra, and allows for the
identification of spectral lines of about 160 ions ranging from He
(Z=2) up to U (Z=92). 
We included neutral and firstly-ionised ions for most of the elements. For some \textit{s}-process elements, 
the second ionisation is
included although the effective temperatures of the two sample stars
in this study are too low for these ionisations.

The equivalent width (EW) of spectral lines are measured interactively with PyMOOG. 
The EWs are calculated with direct integration. The
abundances are computed with an iterative process in which the
theoretical EWs of single lines are computed for given abundances and
matched to the observational EWs. For our analysis, we avoided blended
lines as much as possible. Synthetic spectra were used for checking
whether used spectral lines are part of blends with other identified
lines. 

\subsection{Atmospheric parameters}\label{subsect:atmos}
The atmospheric parameters are determined using the atmospheric
parameter determination module in PyMOOG. We use linear interpolation
to calculate atmospheric models, which are within the parameter steps of
the Kurucz-Castelli models. For the determination of the effective
temperature and surface gravity $\log g$, we use parameter steps of 125 K
and 0.5 dex, respectively. For the microturbulent velocity, we choose
steps of 0.2 km/s.

To derive the atmospheric parameters, we use Fe I and Fe II lines. 
The effective temperature $T_{\textrm{eff}}$ is
determined by enforcing the iron abundance, derived from the
individual Fe I lines, to be independent of lower excitation
potential. We choose Fe I lines for this purpose since the available
Fe II lines do not cover an appropriate range in lower excitation
potential. The surface gravity $\log g$ is determined by enforcing
ionisation equilibrium between the individual Fe I and Fe II
abundances. The microturbulent velocity $\xi_t$ is derived by
enforcing the iron abundance from individual Fe I lines to be
independent of reduced equivalent width, which we define in 
this contribution as EW/$\lambda$.

\begin{table}[tb!]
\caption{\label{table:atmos} Determined atmospheric parameters of J051213 and J051848. The errors for [Fe/H] include line-to-line scatter and model uncertainty. 
Symbols N$_{\textrm{FeI}}$ and N$_{\textrm{FeII}}$ show the number of lines used for Fe I and Fe II, respectively.}
\begin{center}
\begin{tabular}{lcc} \hline\hline
Object  &  J051213  &    J051848   \\   
\hline
T$_{\rm eff}$  (K) & 5875 $\pm$ 125  & 6000 $\pm$ 125 \\
$\log g$ (dex)    & 1.00 $\pm$ 0.25 & 0.50 $\pm$ 0.25 \\
$\xi_t$ (km/s) & 3.0 $\pm$ 0.2   & 2.8 $\pm$ 0.2 \\ 
$\textrm{[FeI/H]}$  &  -0.56 $\pm$ 0.16 &  -1.06 $\pm$ 0.17 \\
$\textrm{[FeII/H]}$ &  -0.56 $\pm$ 0.15 &  -1.03 $\pm$ 0.14 \\
N$_{\textrm{FeI}}$  & 53  &  35 \\
N$_{\textrm{FeII}}$ & 13  &   9 \\
\hline
\end{tabular}
\end{center}
\end{table}

The individual atmospheric parameter results together with the number
of used individual spectral lines are shown in Table \ref{table:atmos}. The
indicated uncertainties of [FeI/H] and [FeII/H] are the total errors on the
iron abundances, which includes line-to-line scatter and the atmospheric
parameter uncertainties as described below in
Sect. \ref{subsect:abun}. The metallicity of J051213 is similar to
the mean LMC metallicity of -0.4 dex, while J051848 has a
significantly lower metallicity.
 
Both stars are of spectral type F and have low surface gravities confirming their evolved nature.
The effective temperatures
of both stars could be determined accurately. 
Kamath et al. (2015,accepted) estimate stellar parameter for J051848, based on low-resolution spectra.
We find that the derived stellar parameters of our study and those of Kamath et al (2015, accepted) agree well 
within the estimated errors.

\subsection{Abundance determination}\label{subsect:abun}

\begin{figure}[t!]
\resizebox{\hsize}{!}{\includegraphics{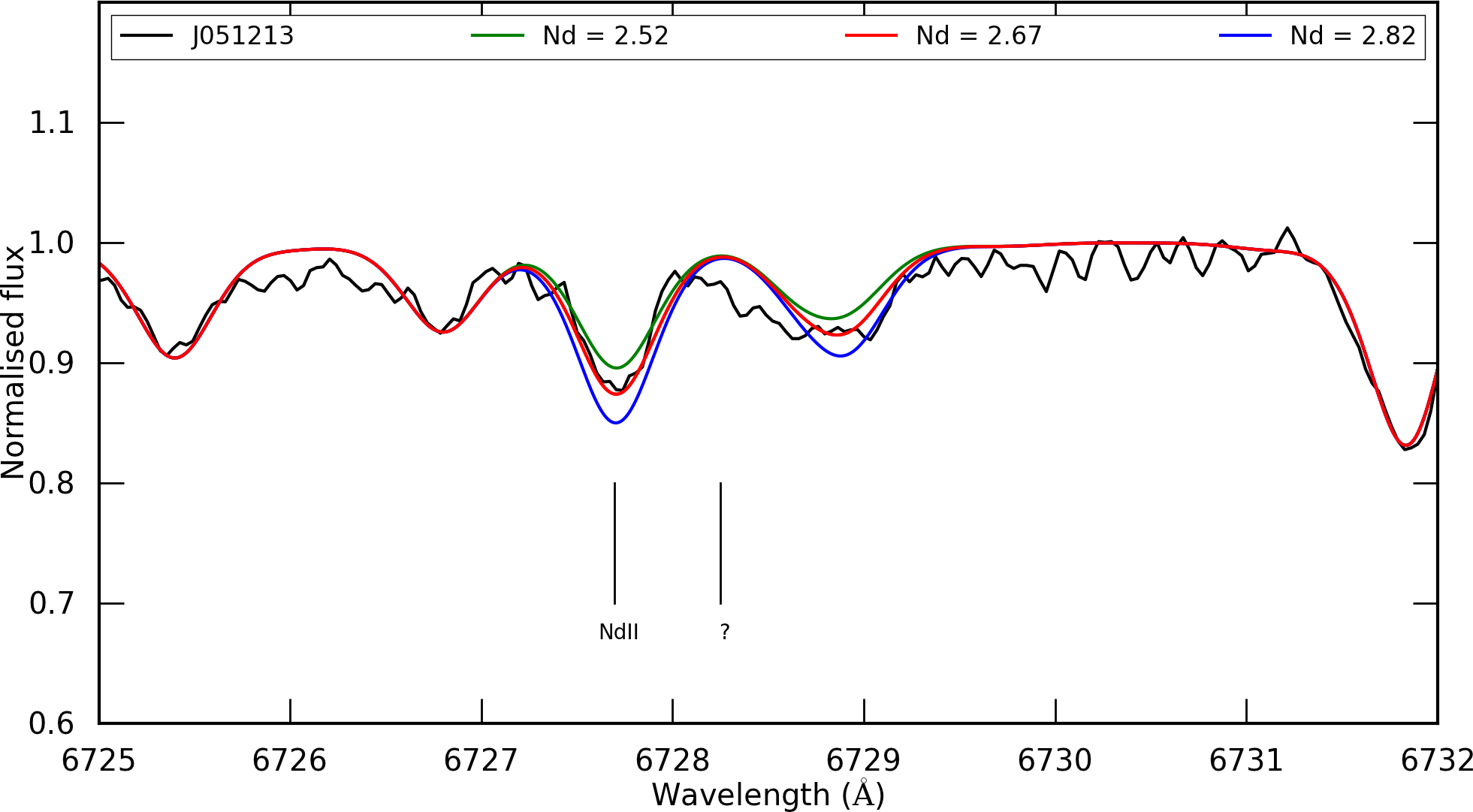}}
\caption{Spectrum synthesis of the Nd II line at 6727.695 \AA{} for J051213. The black spectrum is the observed spectrum,
the coloured spectra represent synthetic spectra with different Th abundances. The position of the Nd line is indicated. 
For more information, see text.}\label{fig:nd}
\end{figure}

\begin{figure}[t!]
\resizebox{\hsize}{!}{\includegraphics{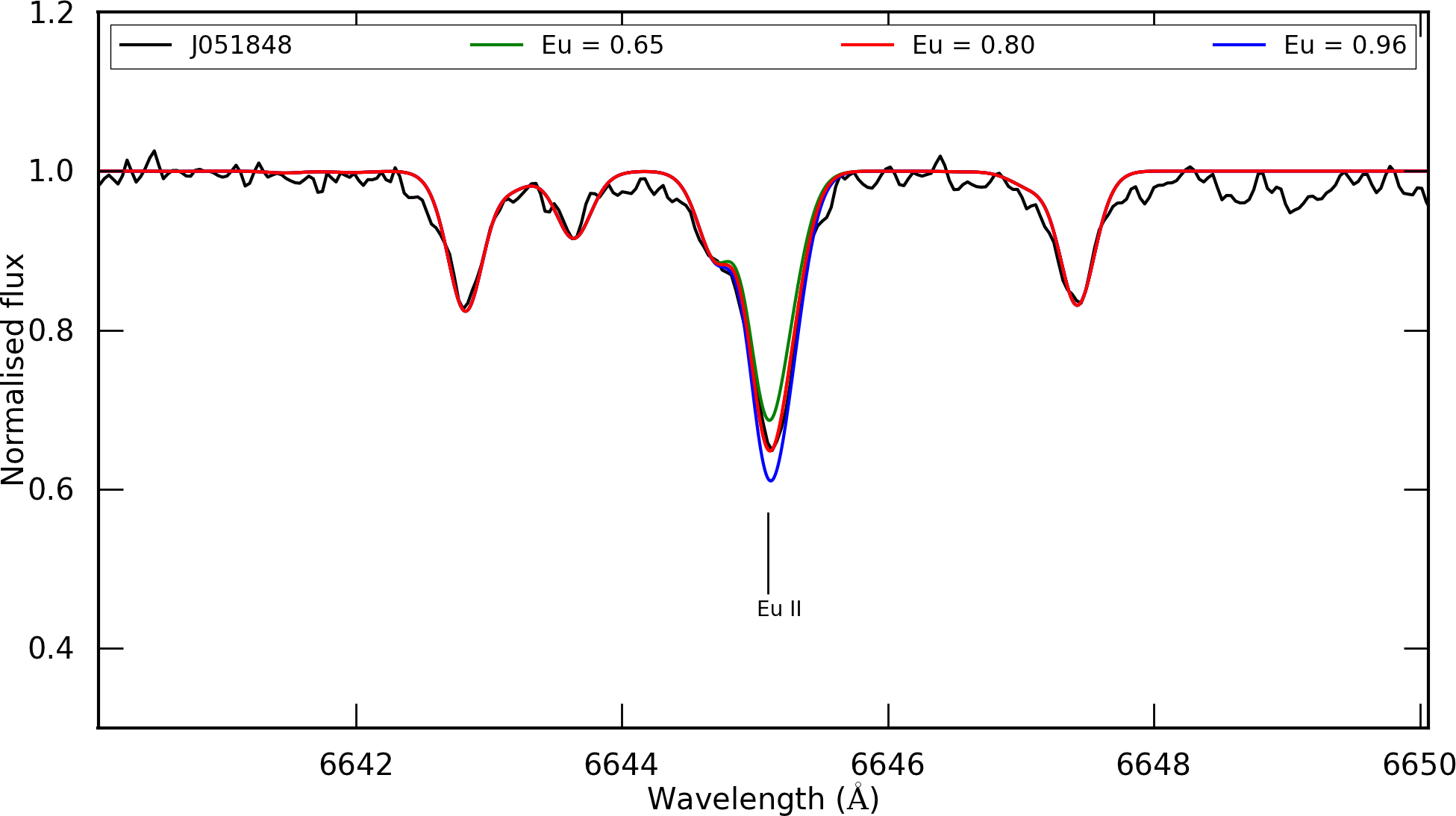}}
\caption{Spectrum synthesis of the Eu II line at 6645.064 \AA{} for J051848. The black spectrum is the observed spectrum,
the coloured spectra represent synthetic spectra with different Th abundances. The position of the Eu line is indicated. 
For more information, see text.}\label{fig:eu}
\end{figure}

We use the derived atmospheric parameters in Table \ref{table:atmos}
for the abundance determination using PyMOOG. 
We mainly use isolated non-blended, non-saturated lines, but this is a challenge because of
the rich spectra with strong enhancements. However,  
with spectral synthesis, we determine the abundances of ions, which can only be found
in blends. Lines with EWs smaller than 5\,m\AA{} are not used as they
may be confused with noise in the spectra.

Only a few carbon and oxygen lines are available for abundance determination because most of the lines are severely blended.
For J051213, where only one spectral
line was available for both the C and O abundances, these abundances were
determined using spectral synthesis. The O abundance of J051213 is
determined using the forbidden [O I] line at 6300\,\AA,{} which is part
of a fully identified blend. This line is not sensitive to non-LTE
effects \citep[e.g.][]{kiselman02}. For J051848, we found four
useful lines for carbon and two non-forbidden O I lines. 
For both stars, nitrogen lines larger than 5\,m\AA{} are not found in the available
wavelength coverage of the UVES spectra.

The spectra of both objects allowed abundance determinations of many
elements past the iron peak. Unfortunately, all available Sr and Ba
lines are heavily saturated hampering accurate abundance
determinations for these two \textit{s}-process elements. Most of the
other studied \textit{s}-elements abundances are determined from multiple,
single lines. Apart from the light \textit{s}-process (ls) peak elements Y
and Zr, and heavy \textit{s}-process (hs) peak elements La, Ce, Pr, and Nd, we
find abundances for heavier elements like gadolinium (Gd, Z=64),
dysprosium (Dy, Z=66), erbium (Er, Z=68), thulium (Tm, Z=69),
luthetium (Lu, Z=71), and hafnium (Hf, Z=72).

To illustrate our analyses, in Fig. \ref{fig:nd} we show the
comparison between synthetic spectra with different Nd abundances for
J051213 around the Nd II line at 6727.695 \AA{}.
The red line represents the best synthetic fit to the
observations. The figure shows that our derived abundances reproduce
well the observed spectral lines.
A similar fit for the Eu II line at 6645.064 \AA{} for J051848 is
shown in Fig. \ref{fig:eu}. 

\begin{figure}[t!]
\resizebox{\hsize}{!}{\includegraphics{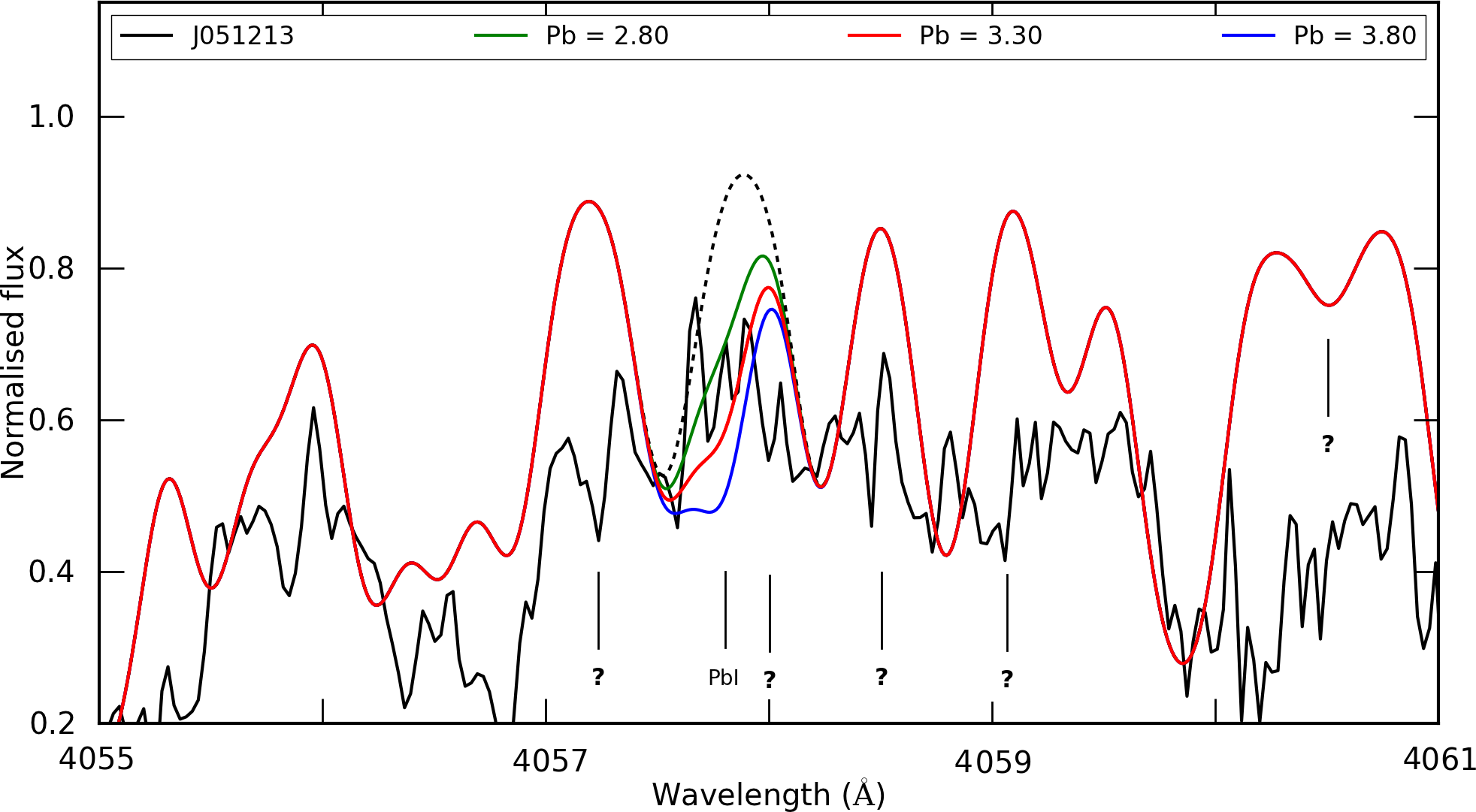}}
\caption{Spectrum synthesis of the Pb I line at 4057.807 \AA{} for J051213. The black spectrum is the observed spectrum of J051213,
the coloured spectra represent synthetic spectra with different Pb abundances. The dashed black line shows the synthetic spectrum if
no Pb is present. Positions of unidentified spectral lines are indicated by question marks. For more information, see text.}\label{fig:Pb_J051213}
\end{figure}

\begin{figure}[t!]
\resizebox{\hsize}{!}{\includegraphics{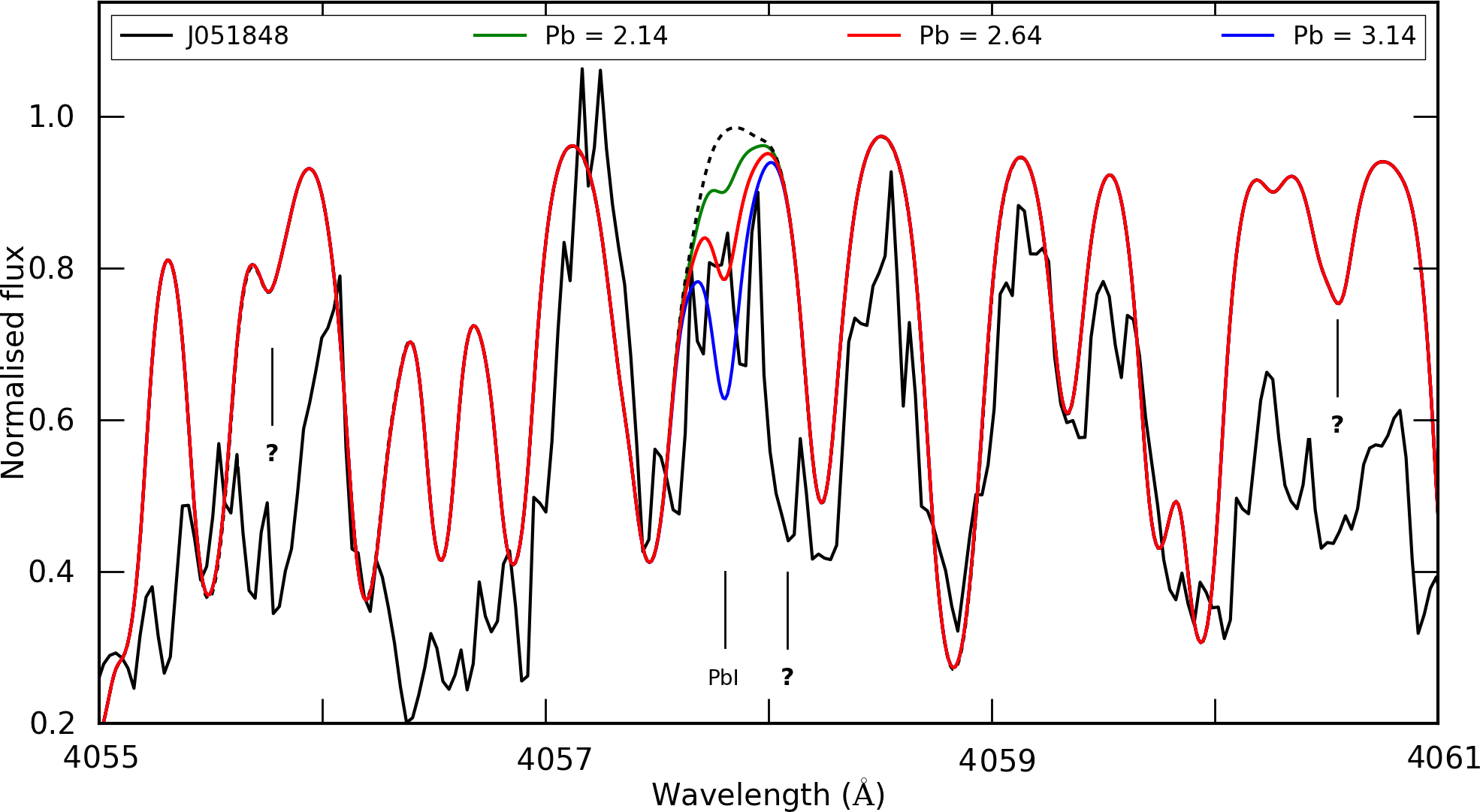}}
\caption{Spectrum synthesis of the Pb I line at 4057.807 \AA{} for J051848. The black spectrum is the observed spectrum,
the coloured spectra represent synthetic spectra with different Pb abundances. The dashed black line shows the synthetic spectrum if
no Pb is present. Positions of unidentified spectral lines are indicated by question marks. For more information, see text.}\label{fig:Pb_J051848}
\end{figure}

Pb is a very important tracer of the \textit{s}-process
nucleosynthesis. Unfortunately, at the photospheric conditions of our
stars, the useful Pb lines are in the very blue part of the spectrum. Because of the low S/N
around the strongest Pb I line at 4057.807\,\AA{}, it is difficult to
determine accurate lead (Pb, Z=82) abundances for both sample
stars, therefore, we prefer to determine Pb abundance upper limits.  
For J051213, the S/N is poor as can be seen in
Fig. \ref{fig:Pb_J051213}. The combination of many unidentified
spectral lines, i.e. lines not included in the line-lists, and the poor
S/N make it difficult to determine the position of the continuum. For
some unidentified lines, we mark the positions indicated with question
marks. 
For J051848, the S/N is higher, but again many unidentified
lines are present in the spectral region (as shown in
Fig. \ref{fig:Pb_J051848}). For both stars, we use the spectral
blended lines at 4057.5, 4058.3, 4058.8, and 4059.9\,\AA{} to estimate the position of the continuum 
at the Pb I line by eye. With our adopted local continuum positions,
small Pb line features are present in the observed spectra as the
dashed lines represent synthetic spectra without the Pb I line at
4057.807\,\AA{}.

\begin{table}[t!]
\begin{threeparttable}
\caption{\label{table:abun_J051213} Abundance results of J051213.81-693537.1.}
\setlength{\tabcolsep}{0.15cm}
\begin{tabular}{l | ccccc | c } \hline\hline
    & \multicolumn{5}{c}{J051213} & \multicolumn{1}{|c}{Sun} \\
    & \multicolumn{2}{c}{$T_{\textrm{eff}}$ = 5875 K} & & \multicolumn{2}{c}{$\xi_t$ = 3.0 km/s} & \multicolumn{1}{|c}{}\\
    & \multicolumn{2}{c}{$\log g$ = 1.00 dex} & & \multicolumn{2}{c}{[Fe/H] = -0.56 dex} & \multicolumn{1}{|c}{}\\
\hline    
Ion & N  &  log $\epsilon$  &  $\sigma_{\textrm{l2l}}$  &  [X/Fe] &  $\sigma_{tot}$  & log $\epsilon_{\odot}$ \\
\hline
C I & 1 & 8.75 & 0.20 & 0.88 & 0.26 & 8.43 \\
O I & 1 & 8.65 & 0.20 & 0.52 & 0.26 & 8.69 \\
\hline
Mg I & 1 & 7.00 & 0.20 & -0.04 & 0.24 & 7.60 \\
Al I & 1 & 6.60 & 0.20 & 0.71 & 0.25 & 6.45 \\
Si I & 4 & 7.32 & 0.05 & 0.37 & 0.18 & 7.51 \\
S I & 2 & 6.51 & 0.03 & -0.05 & 0.13 & 7.12 \\
Ca I & 11 & 6.00 & 0.09 & 0.22 & 0.09 & 6.34 \\
Ca II & 1 & 6.00 & 0.20 & 0.22 & 0.24 & 6.34 \\
Sc II & 3 & 2.93 & 0.03 & 0.34 & 0.10 & 3.15 \\
Ti I & 2 & 4.69 & 0.07 & 0.30 & 0.13 & 4.95 \\
Ti II & 7 & 4.62 & 0.13 & 0.23 & 0.11 & 4.95 \\
V I & 1 & 3.91 & 0.20 & 0.54 & 0.25 & 3.93 \\
V II & 2 & 3.91 & 0.04 & 0.54 & 0.14 & 3.93 \\
Cr I & 4 & 5.20 & 0.07 & 0.12 & 0.10 & 5.64 \\
Cr II & 7 & 5.16 & 0.11 & 0.08 & 0.10 & 5.64 \\
Mn I & 1 & 4.69 & 0.20 & -0.18 & 0.24 & 5.43 \\
Fe I & 53 & 6.94 & 0.10 & 0.00 & 0.06 & 7.50 \\
Fe II & 13 & 6.94 & 0.09 & 0.00 & 0.07 & 7.50 \\
Ni I & 4 & 5.75 & 0.08 & 0.09 & 0.12 & 6.22 \\
Zn I & 1 & 4.22 & 0.20 & 0.22 & 0.29 & 4.56 \\
\hline
Y II & 5 & 3.13 & 0.11 & 1.48 & 0.13 & 2.21 \\
Zr II & 2 & 3.20 & 0.02 & 1.18 & 0.10 & 2.58 \\
\hline
La II & 6 & 2.52 & 0.13 & 1.98 & 0.13 & 1.10 \\
Ce II & 6 & 2.91 & 0.12 & 1.89 & 0.13 & 1.58 \\
Pr II & 6 & 1.96 & 0.06 & 1.80 & 0.13 & 0.72 \\
Nd II & 14 & 2.67 & 0.10 & 1.81 & 0.13 & 1.42 \\
Sm II & 1 & 1.70 & 0.20 & 1.30 & 0.24 & 0.96 \\
Eu II & 2 & 0.98 & 0.07 & 1.02 & 0.13 & 0.52 \\
\hline
Dy II & 2 & 2.00 & 0.20 & 1.46 & 0.21 & 1.10 \\
Er II & 2 & 1.90 & 0.20 & 1.54 & 0.22 & 0.92 \\
Pb I$^{\textrm{u}}$ & 1 & 3.30 & 0.20 & 2.11 & 0.25 & 1.75 \\
Th II$^{\textrm{u}}$ & 1 & 1.35 & 0.20 & 1.89 & 0.27 & 0.02 \\
\hline
\end{tabular}
    \begin{tablenotes}
      \small
      \item \textbf{Notes:} The table lists for each ion the used number of lines (N) for the abundance determination, the determined abundance (log $\epsilon$)
      , the uncertainty of this 
      abundance due to line-to-line scatter ($\sigma_{\textrm{l2l}}$), the element over iron ratio ([X/Fe]), and total uncertainty ($\sigma_{tot}$). 
      The total uncertainty $\sigma_{tot}$ includes line-to-line scatter and 
      atmospheric parameter uncertainty (see text for details). The last column lists the solar abundances from \citet{asplund09}.
      \item We choose a standard deviation of 0.20 dex for all ions for which the abundance was derived using spectrum synthesis or for which only one line is available.
      \item $^{(\textrm{u})}$ These abundances are upper limits.
    \end{tablenotes}
 \end{threeparttable}
\end{table}

\begin{table}[t!]
\begin{threeparttable}
\caption{\label{table:abun_J051848} Abundance results for J051848.86-700246.9.}
\setlength{\tabcolsep}{0.15cm}
\begin{tabular}{l | ccccc | c } \hline\hline
    & \multicolumn{5}{c}{J051848} & \multicolumn{1}{|c}{Sun} \\
    & \multicolumn{2}{c}{$T_{\textrm{eff}}$ = 6000 K} & & \multicolumn{2}{c}{$\xi_t$ = 2.8 km/s} & \multicolumn{1}{|c}{}\\
    & \multicolumn{2}{c}{$\log g$ = 0.50 dex} & & \multicolumn{2}{c}{[Fe/H] = -1.03 dex} & \multicolumn{1}{|c}{}\\    
\hline    
Ion & N  &  log $\epsilon$  &  $\sigma_{\textrm{l2l}}$  &  [X/Fe] &  $\sigma_{tot}$  & log $\epsilon_{\odot}$ \\
\hline
C I & 4 & 8.61 & 0.13 & 1.21 & 0.16 & 8.43 \\
O I & 2 & 8.50 & 0.05 & 0.84 & 0.19 & 8.69 \\
\hline
Mg I & 2 & 6.73 & 0.07 & 0.19 & 0.12 & 7.60 \\
S I & 1 & 6.46 & 0.20 & 0.37 & 0.25 & 7.12 \\
Ca I & 8 & 5.56 & 0.08 & 0.28 & 0.08 & 6.34 \\
Sc II & 6 & 2.30 & 0.08 & 0.18 & 0.11 & 3.15 \\
Ti II & 4 & 4.12 & 0.08 & 0.20 & 0.11 & 4.95 \\
Cr I & 6 & 4.70 & 0.07 & 0.12 & 0.08 & 5.64 \\
Cr II & 12 & 4.78 & 0.07 & 0.17 & 0.08 & 5.64 \\
Fe I & 35 & 6.44 & 0.07 & 0.00 & 0.05 & 7.50 \\
Fe II & 9 & 6.47 & 0.12 & 0.00 & 0.10 & 7.50 \\
Ni I & 8 & 5.22 & 0.10 & 0.06 & 0.10 & 6.22 \\
Cu I & 1 & 3.53 & 0.20 & 0.40 & 0.23 & 4.19 \\
Zn I & 1 & 3.78 & 0.20 & 0.25 & 0.29 & 4.56 \\
\hline
Y II & 6 & 2.79 & 0.12 & 1.61 & 0.15 & 2.21 \\
Zr II & 2 & 2.85 & 0.05 & 1.30 & 0.13 & 2.58 \\
\hline
La II & 7 & 2.55 & 0.11 & 2.48 & 0.20 & 1.10 \\
Ce II & 5 & 2.37 & 0.07 & 1.82 & 0.13 & 1.58 \\
Pr II & 11 & 1.75 & 0.05 & 2.06 & 0.14 & 0.72 \\
Nd II & 14 & 2.69 & 0.10 & 2.30 & 0.16 & 1.42 \\
Sm II & 2 & 1.80 & 0.12 & 1.87 & 0.18 & 0.96 \\
Eu II & 2 & 0.81 & 0.12 & 1.32 & 0.17 & 0.52 \\
\hline
Gd II & 10 & 1.75 & 0.12 & 1.71 & 0.12 & 1.07 \\
Dy II & 4 & 1.91 & 0.09 & 1.84 & 0.12 & 1.10 \\
Er II & 1 & 1.63 & 0.20 & 1.74 & 0.26 & 0.92 \\
Tm II & 1 & 1.20 & 0.20 & 2.13 & 0.24 & 0.10 \\
Lu II & 3 & 1.16 & 0.08 & 2.08 & 0.15 & 0.10 \\
Hf II & 1 & 1.84 & 0.20 & 2.02 & 0.25 & 0.85 \\
Pb I & 1 & 2.62 & 0.20 & 1.93 & 0.25 & 1.75 \\
\hline
\end{tabular}
    \begin{tablenotes}
      \small
      \item \textbf{Notes:} Same as for Table \ref{table:abun_J051213}. 
    \end{tablenotes}
\end{threeparttable}
\end{table}

\section{Abundance results}\label{sect:abun}

The complete abundance analysis results of both objects are presented in Tables \ref{table:abun_J051213} and \ref{table:abun_J051848}. 
An overview of the lines used for these analyses can be found in two catalogues, which are available at CDS\footnote{A link to the catalogues}. 
The information in each catalogue is tabulated as follows: Col.1 lists the name of the ion to which the corresponding line belongs, Col. 2 shows the 
rest wavelength, Col. 3 contains the lower excitation potential, Col. 4 mentions the logarithm of the  oscillator strength, Col. 5 lists the measured EW, 
 and Col. 6 gives the deduced abundance for the corresponding line.

\begin{figure}[t!]
\resizebox{\hsize}{!}{\includegraphics{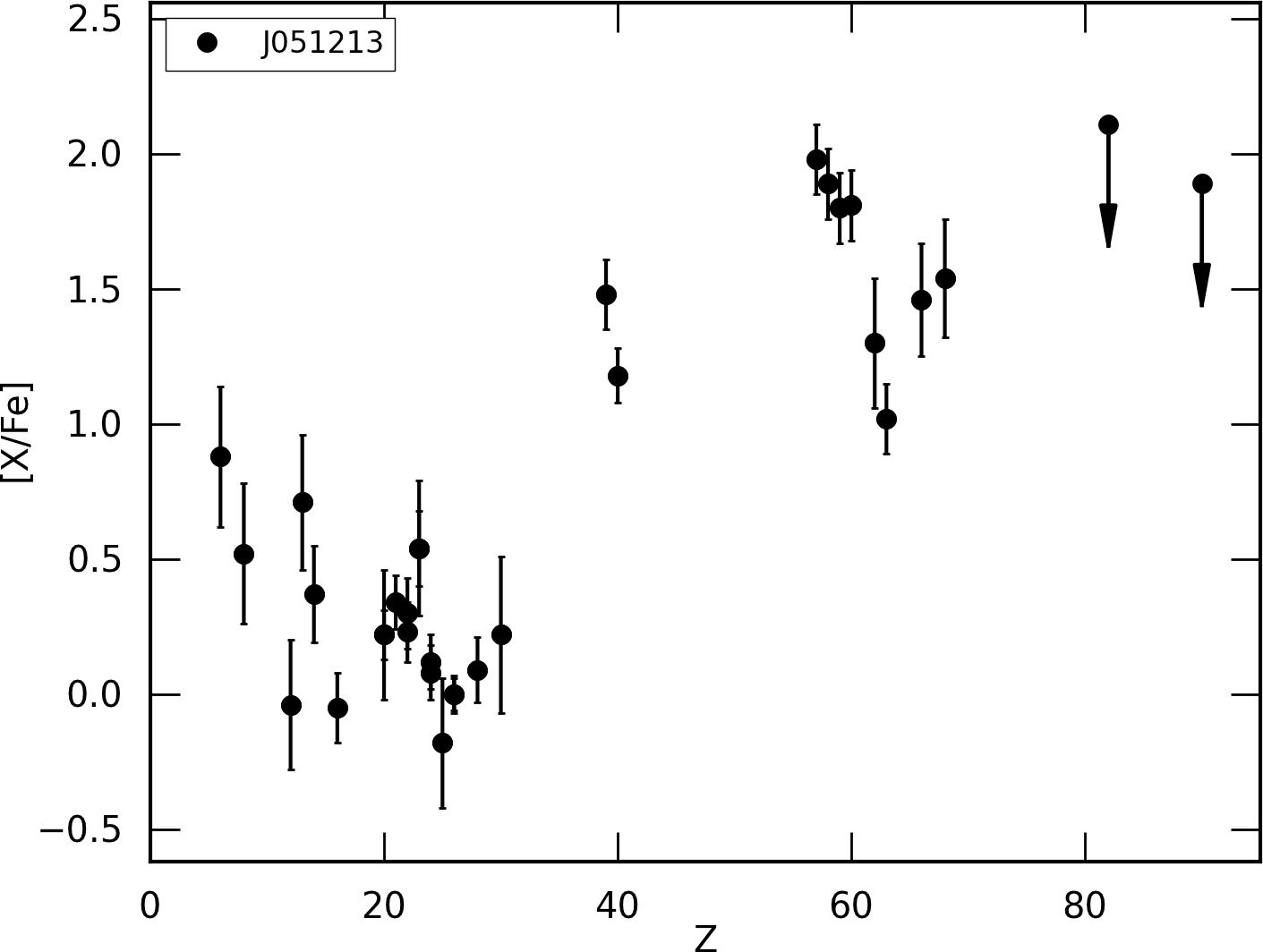}}
\caption{[X/Fe] results of J051213. The errors bars represent the total uncertainties $\sigma_{\textrm{tot}}$. The abundances of Pb 
(Z=82) and Th (Z=90) are upper limits and marked with down arrows.}\label{fig:abun_J051213}
\end{figure}

\begin{figure}[t!]
\resizebox{\hsize}{!}{\includegraphics{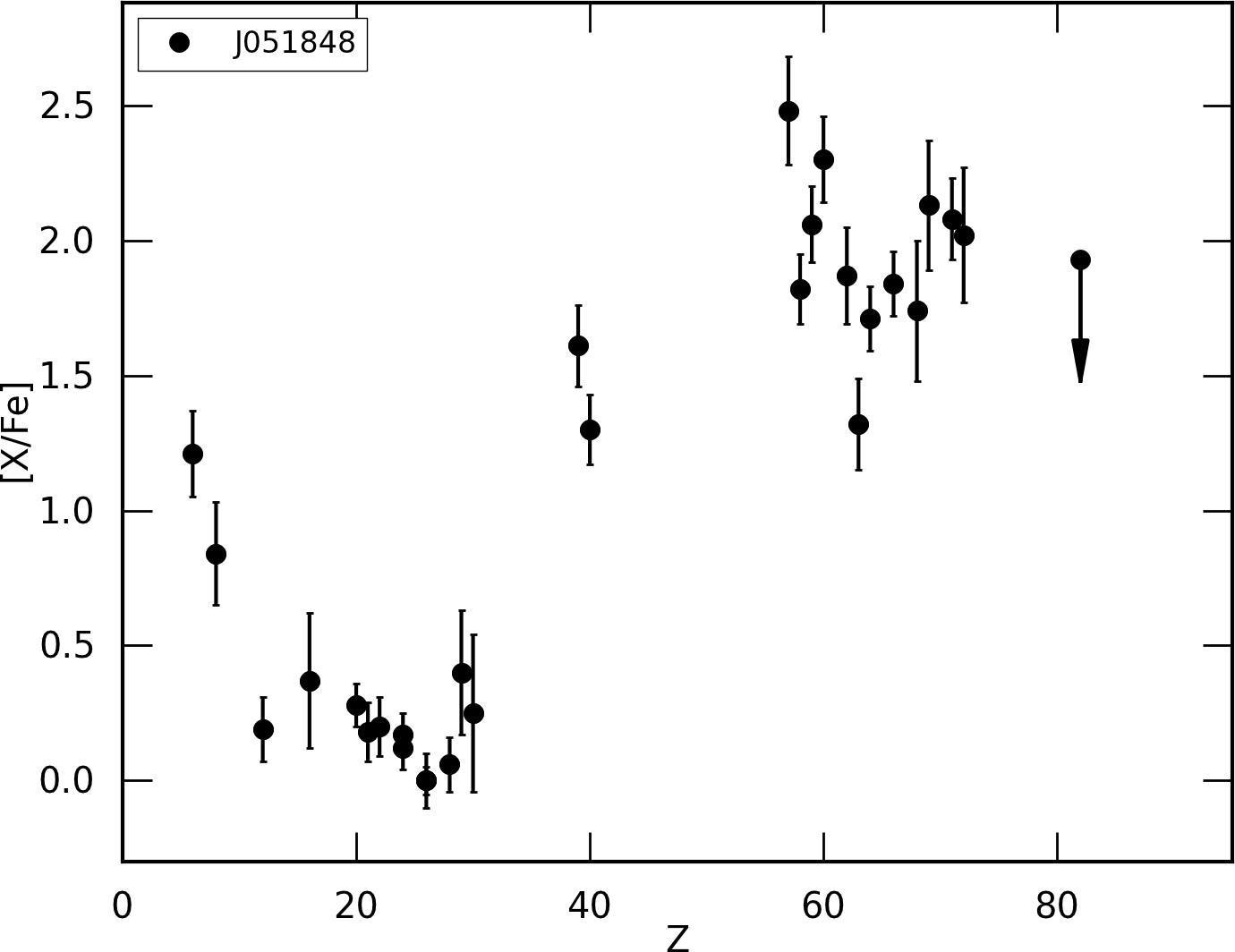}}
\caption{[X/Fe] results of J051848. The errors bars represent the total uncertainties $\sigma_{\textrm{tot}}$. The abundance of Pb 
(Z=82) is an upper limit and marked with a down arrow.}\label{fig:abun_J051848}
\end{figure}

Based upon the ionisation potential of the corresponding ion, the element over iron ratios ([X/Fe]) in Tables \ref{table:abun_J051213} and 
\ref{table:abun_J051848} are calculated using Fe I or Fe II. If the ionisation potential of an ion is below the ionisation potential of Fe I, 
the abundance of Fe I is used for calculating [X/Fe]. If the ionisation potential exceeds the ionisation potential of Fe I, Fe II is used for 
calculating [X/Fe]. The same principle is also used for calculating the total error of [X/Fe]. The errors were determined using the method 
described in \citet{deroo05a}. The uncertainties due to the atmospheric parameters are calculated by determining the abundances of a certain ion 
for atmospheric models with $T_{\textrm{eff}}$ plus and minus 125 K, models with $\log g$ plus and minus 0.25 dex and microturbulent velocity plus 
and minus 0.2 km/s (see Table \ref{table:atmos}). The uncertainty in microturbulent velocity can have important effects for ions, which are 
only determined by large lines like Y II. The total 
uncertainties are then the quadratic sum of the uncertainties of the mean due to line-to-line scatter ($\sigma_{\textrm{l2l}}$), uncertainties due to 
atmospheric parameters ($\sigma_{\textrm{T}_{eff}}$, $\sigma_{\textrm{logg}}$, $\sigma_{\xi_t}$), and the Fe abundance uncertainty ($\sigma_{\textrm{Fe}}$):
\begin{equation*}
\sigma_{\textrm{tot}} = \sqrt{\left(\frac{\sigma_{\textrm{l2l}}}{\sqrt{N_{\textrm{ion}}}}\right)^2 + (\sigma_{\textrm{T}_{eff}})^2 
+ (\sigma_{\textrm{logg}})^2 + (\sigma_{\xi_t})^2  + \left(\frac{\sigma_{\textrm{Fe}}}{\sqrt{N_{\textrm{Fe}}}}\right)^2} .
\end{equation*}
The scaling to Fe I or Fe II based upon ionisation potential strongly decreases the [X/Fe] uncertainty of certain ions with respect to using 
only Fe I or only Fe II for all [X/Fe] calculations.
The [X/Fe] results in Tables \ref{table:abun_J051213} and \ref{table:abun_J051848} are plotted in Figs. \ref{fig:abun_J051213} and \ref{fig:abun_J051848}, respectively.

\subsection{Carbon and oxygen}
Both stars are carbon enhanced. The uncertainties of C and O for
J051213 are large since only one line could be used for the abundance
determinations. Both ions are also strongly sensitive to changes in
effective temperature. We find [C/Fe] values of about 0.9 and 1.2
for J051213 and J051848,  and [O/Fe] values
of about 0.5 and 0.8. This
results in moderate C/O ratios of only 1.26 $\pm$ 0.40 for J051213 and 1.29
$\pm$ 0.30 for J051848, which are a combination of relatively low C enhancements
as well as relatively high O enhancements.  Based upon their abundances (Fig. \ref{fig:abun_J051213} 
and \ref{fig:abun_J051848}) and their luminosities
(see Sect. \ref{sect:sed}), both stars are carbon-enhanced post-AGB stars but the C/O
ratios are not very large.

\subsection{$ \alpha$-elements}  
Concerning the available $\alpha$-elements Mg, Si, S, Ca, and Ti, the
mean of [X/Fe] is [$\alpha$/Fe] = +0.14 and [$\alpha$/Fe] =
+0.26 for J051213 and J051848, respectively. For J051848, no Si
abundance was available. With respect to Galactic abundances, these
$\alpha$-element abundance are slightly deficient but they fall within
the expected abundance ranges for the LMC, consistent with their
respective metallicities \citep[e.g][]{vanderswaelmen13,pompeia08}.

\subsection{\textit{s}-process elements}

The \textit{s}-process elements can be subdivided into three groups. This division is based
upon the number of neutrons in the nuclei. The first group is  the
light \textit{s}-process (ls) elements around neutron magic number 50 (Z $\sim$ 38) with
elements like Sr, Y, and Zr. For both stars, we were only able to measure Y and Zr.
The second group is the heavy \textit{s}-process
(hs) elements around magic neutron number 82 (Z $\sim$ 58) 
with elements like Ba,
La, Ce, Pr, Nd, and Sm. The last group consists of one element,
which is the double magic ion $^{208}$Pb. This has a magic neutron
number of 126 and a magic proton number of 82. Because of its double magic
state, $^{208}$Pb is accepted as the end product of \textit{s}-process
nucleosynthesis.

The [X/Fe] results in Figs. \ref{fig:abun_J051213} and
\ref{fig:abun_J051848} show strong enrichment in both stars,
confirming their post third TDU status. In particular, J051848
shows strong enhancements for La (Z=57) and Nd (Z=60), two typical
\textit{s}-process elements of Ba peak. Also, the \textit{s}-elements of the Sr peak are
enriched, albeit less strongly then the Ba-peak elements. The
abundances of elements past the hs peak are strongly enhanced in both
stars.

For J051213, we find a Pb abundance upper limit  that is similar to the
overabundances of the Ba-peak elements. For
J051848, we find that the Pb upper abundance limit is lower than
the abundance ratios of La and Nd, and similar to the Ce (Z=58)
abundance ratio. 

An overview of the C/O ratio, metallicity, $\alpha$-element enhancement, and the observational indices used 
for describing \textit{s}-process overabundances and distributions (See Sect. \ref{sect:neutron}) are listed in Table \ref{table:ratios}. 

\begin{table*}[t!]
\caption{\label{table:ratios} Overview of the C/O ratio, metallicity, $\alpha$-element enrichment, and \textit{s}-process indices for J051213 and J051848.}
\begin{center}
\begin{tabular}{lccccccc} \hline\hline
Object  &  C/O  &    [Fe/H] &  [$\alpha$/Fe]  &  [ls/Fe]  &  [hs/Fe]  &  [s/Fe]  &  [hs/ls]  \\   
\hline
J051213  &  1.26 $\pm$ 0.40  &  -0.56  $\pm$ 0.15  & 0.14 $\pm$ 0.07 &  1.33 $\pm$ 0.08 & 1.74 $\pm$ 0.08 & 1.61 $\pm$ 0.06 & 0.41 $\pm$ 0.12  \\
J051848  &  1.29 $\pm$ 0.30  &  -1.03  $\pm$ 0.14  & 0.26 $\pm$ 0.08 & 1.46 $\pm$ 0.10 & 2.12 $\pm$ 0.08 & 1.90 $\pm$ 0.07 & 0.66 $\pm$ 0.13  \\
\hline
\end{tabular}
\end{center}
\end{table*}

\section{Luminosity and initial mass determination}\label{sect:sed}

\subsection{Spectral energy distributions and luminosities}\label{subsect:sed}
\begin{table}[tb!]
\caption{\label{table:sed} Estimated luminosity and reddening of J051213 and J051848.}
\begin{center}
\begin{tabular}{lcc} \hline\hline
Object  &  J051213  &    J051848   \\   
\hline
E(B-V) & 0.69 $\pm$ 0.03  & 0.44 $\pm$ 0.02 \\
$L$\,(L$_{\odot}$)  & 6700 $\pm$ 200 & 6250 $\pm$ 200 \\ 
\hline
\end{tabular}
\end{center}
\end{table}

The spectral energy distributions (SEDs) and known distance to the LMC
allow us to determine the luminosities of our sample stars and
estimate their initial masses (Sect. \ref{subsect:inimass}). The
photometric data for constructing the SEDs are retrieved from the
following catalogues: the UBVR
CCD Survey of the Magellanic Clouds \citep{massey02}, 
the 2MASS 6X catalogue \citep{cutri03},
the Deep Near-Infrared Survey (DENIS,\citet{fouque00}), 
the WISE All-Sky
Data catalogue \citep{cutri12}, and 
the Spitzer SAGE survey of the LMC
\citep{meixner06}.

For our sample post-AGB stars, we assume that three sources  contribute to the reddening. 
The first source is reddening by interstellar dust in the
Milky Way galaxy towards the LMC. \citet{schlegel98} find a relatively
small extinction towards the LMC of E(B-V) = 0.075 mag. The second
source is reddening by interstellar dust in the LMC. We assume similar
extinction laws in the LMC and Milky Way galaxy and use the Galactic
extinction curves of \citet{cardelli89} to determine reddening in the
LMC. The third source of reddening is caused by circumstellar dust of the
post-AGB object itself. We assume the circumstellar extinction has the same 
wavelength dependency as the Galactic extinction law.

\begin{figure}[t!]
\resizebox{\hsize}{!}{\includegraphics{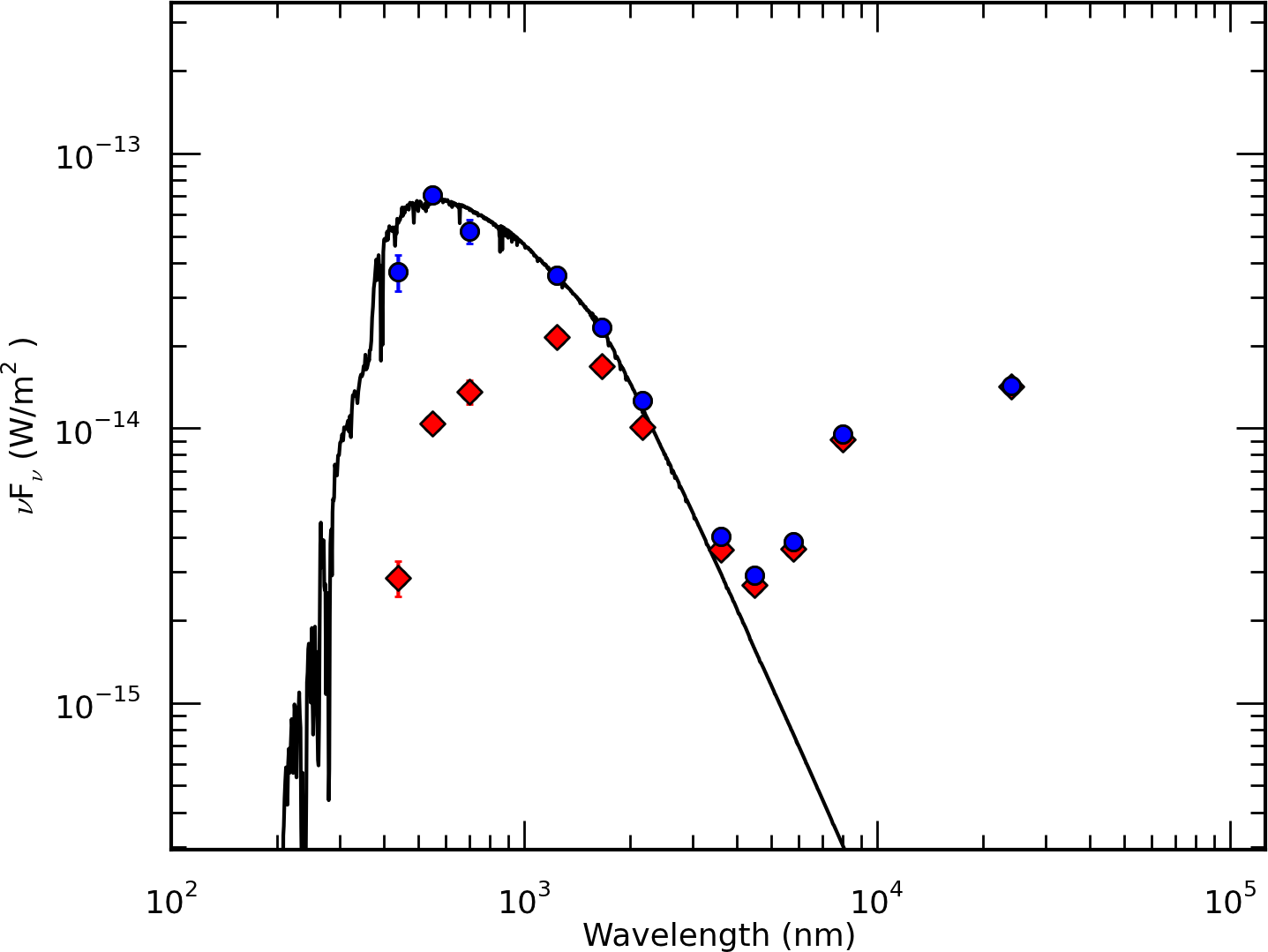}}
\caption{SED of J051213. Red dots symbolise the red, original photometry, blue dots represent the dereddened photometry. 
The black line is the scaled Kurucz atmosphere model.}\label{fig:sed_J051213}
\end{figure}

\begin{figure}[t!]
\resizebox{\hsize}{!}{\includegraphics{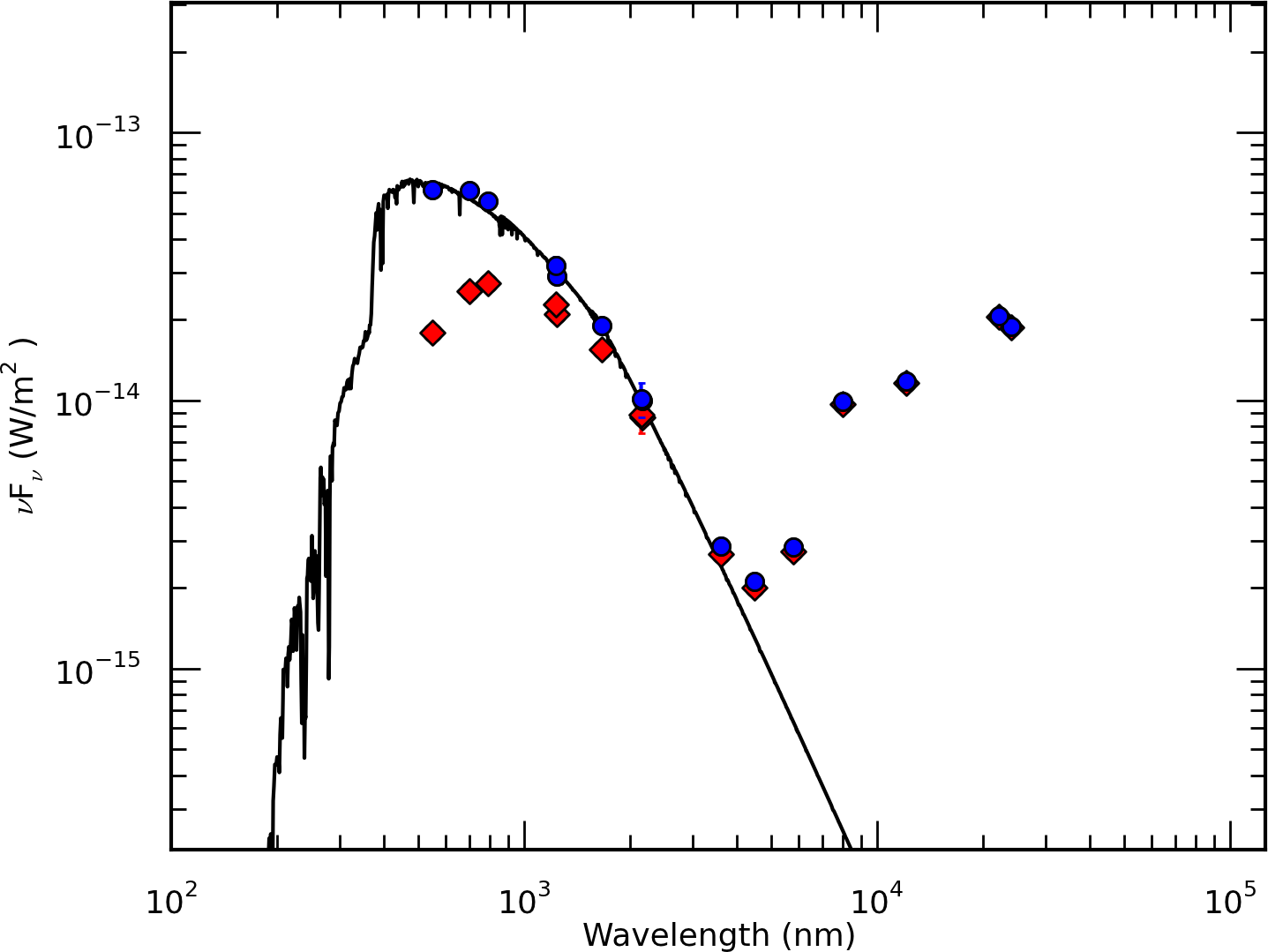}}
\caption{SED of J051848. Symbols are identical to Fig. \ref{fig:sed_J051213}.}\label{fig:sed_J051848}
\end{figure}

The total dereddening is determined by applying a $\chi^2$
minimalisation on the fit between the dereddened broadband fluxes and
the appropriate Kurucz model atmospheres that are used for the abundance
determination. For the scaled model of J051213, we have linearly
interpolated between available models to obtain our preferred
atmosphere model. The error on E(B-V) is determined by a Monte Carlo
simulation of 100 arrays with a normal distribution of the original
flux. The luminosities are calculated by integrating the surface of
the scaled Kurucz models and applying a distance of 50 kpc to the
LMC \citep[e.g.][]{keller06,reid10}. The errors on the luminosity are determined by a Monte Carlo
simulation, similar to the error derivation of the reddening. The estimated luminosity and reddening
 results of both the sample stars are listed in Table
\ref{table:sed}. The constructed SEDs are shown in
Figs. \ref{fig:sed_J051213} and \ref{fig:sed_J051848}.

The SEDs in Figs. \ref{fig:sed_J051213} and \ref{fig:sed_J051848} are
double peaked, indicating a circumstellar dust envelope originating
from the strong mas\textit{s}-loss during the AGB \citep{vanwinckel03}. We find a significant reddening
value for both stars; in particular, J051213 has a high line-of-sight extinction of
E(B-V) $\approx$ 0.7. This extinction is dominated by the circumstellar
extinction given the evidence for dusty circumstellar envelopes of these objects. 
The derived
luminosities in Table \ref{table:sed} are within the expected
luminosity range for post-AGB stars and they agree well with the post-TDU nature of these 
objects.

\subsection{Initial mass estimates}\label{subsect:inimass}

\begin{figure}[t!]
\resizebox{\hsize}{!}{\includegraphics{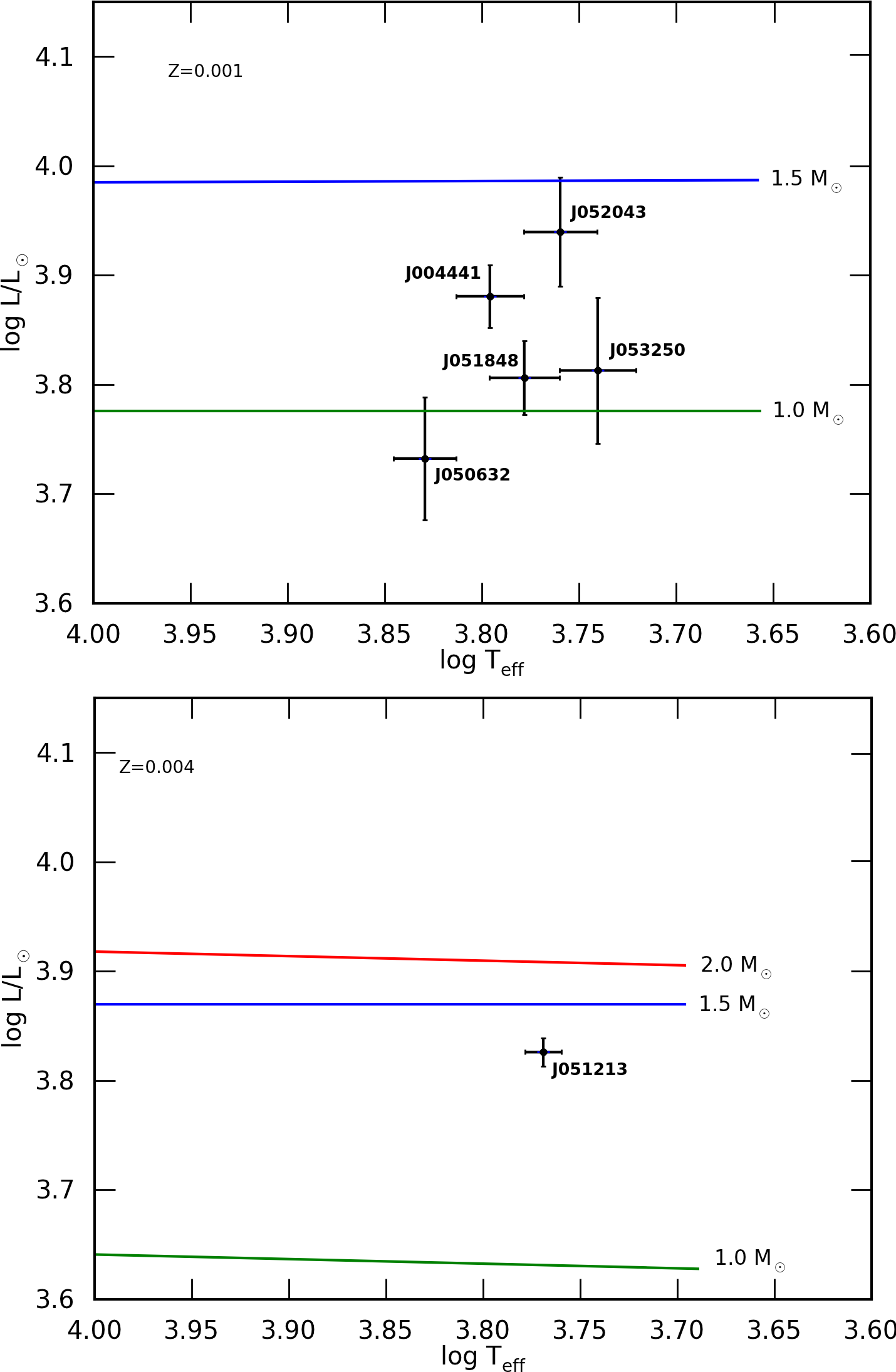}}
\caption{Comparison of our sample stars and four other \textit{s}-process enriched Magellanic Cloud post-AGB stars. The  evolutionary tracks shown are from 
\citet{vassiliadis94} with a metallicity of Z=0.001 in the upper panel and Z=0.004 in the lower panel. The mass corresponding to each track is indicated. 
For more information, see text.}\label{fig:inimass}
\end{figure}

Two key parameters in state-of-the-art evolution and nucleosynthetic AGB models are
the metallicity and initial stellar mass. We use the 
post-AGB evolutionary tracks of \citet{vassiliadis94} to obtain an estimate of the
initial masses of our sample stars. We compare the positions of the
sample stars with those of the theoretical tracks in the
Hertsprung-Russell (HR) diagram in Fig. \ref{fig:inimass}. We have
also included the positions of three other \textit{s}-process enriched LMC
post-AGB stars: J050632.10-714229.9, J052043.86-692341.0, and
J053250.69-713925.8 \citep{vanaarle13}, and the \textit{s}-process enriched SMC
post-AGB star J004441.04-732136.4 \citep{desmedt12}. For all stars, we
use the theoretical tracks that were calculated for  metallicity,
which corresponded the most to the derived metallicities of the individual objects. 
The
theoretical tracks from \citet{vassiliadis94} start at an effective
temperature of 10$^4$ K, which is higher than the derived temperatures
of the shown stars. Since the luminosity of post-AGB stars remains
approximately constant when crossing the Hr-diagram, we have used
linear extrapolation between log $T_{\textrm{eff}}$ and log
$L$/$L_{\odot}$ to extrapolate towards lower temperatures.

We find that the \textit{s}-process enriched post-AGB stars included in
Fig. \ref{fig:inimass} have low masses below about 1.5
M$_{\odot}$. We find that strong \textit{s}-process enrichment is linked to 
low initial masses in all sources studied so far  \citep[][]{desmedt12,vanaarle13}.

We remark that our initial mass estimates depend on the applied mass-loss rate history on the
AGB, and the theoretical post-AGB tracks of \citet{vassiliadis94} used for the initial mass estimates, are 
determined using  initial-final mass relations deduced at that time.

\begin{table*}[t!]
\caption{\label{table:sindex} Overview of the four \textit{s}-process indices of the stars in Figs. \ref{fig:sfe} and \ref{fig:feh}.}
\begin{threeparttable}
\begin{tabular}{lcccccc} \hline\hline
Object  &  [Fe/H] & C/O & [ls/Fe]  &  [hs/Fe]  &  [s/Fe]  &  [hs/ls]  \\   
\hline
IRAS 04296+3429$^{a}$ & -0.62 $\pm$ 0.20 &               & 1.66 $\pm$ 0.12 & 1.34 $\pm$ 0.07 & 1.45 $\pm$ 0.06 & -0.32 $\pm$ 0.14 \\  
IRAS 05341+0852$^{a}$ & -0.85 $\pm$ 0.20 & 1.4 $\pm$ 0.3 & 1.92 $\pm$ 0.08 & 2.30 $\pm$ 0.07 & 2.17 $\pm$ 0.05 & 0.38 $\pm$ 0.11 \\ 
IRAS 06530-0213$^{b}$ & -0.46 $\pm$ 0.20 & 2.8 $\pm$ 0.3 & 1.82 $\pm$ 0.09 & 2.11 $\pm$ 0.07 & 2.02 $\pm$ 0.05 & 0.29 $\pm$ 0.11 \\   
IRAS 07134+1005$^{a}$ & -1.00 $\pm$ 0.20 & 1.0 $\pm$ 0.3 & 1.55 $\pm$ 0.17 & 1.36 $\pm$ 0.07 & 1.43 $\pm$ 0.07 & -0.19 $\pm$ 0.18 \\
IRAS 08143-4406$^{b}$ & -0.39 $\pm$ 0.20 & 1.3 $\pm$ 0.3 & 1.52 $\pm$ 0.15 & 1.49 $\pm$ 0.07 & 1.50 $\pm$ 0.07 & -0.03 $\pm$ 0.16 \\
IRAS 19500-1709$^{a}$ & -0.60 $\pm$ 0.20 & 1.1 $\pm$ 0.3 & 1.38 $\pm$ 0.11 & 0.90 $\pm$ 0.07 & 1.06 $\pm$ 0.06 & -0.48 $\pm$ 0.13 \\
IRAS 22223+4327$^{a}$ & -0.31 $\pm$ 0.20 & 1.2 $\pm$ 0.3 & 1.55 $\pm$ 0.10 & 1.25 $\pm$ 0.09 & 1.35 $\pm$ 0.07 & -0.30 $\pm$ 0.14 \\ 
IRAS 23304+6147$^{a}$ & -0.79 $\pm$ 0.20 & 2.9 $\pm$ 0.3 & 1.55 $\pm$ 0.10 & 1.50 $\pm$ 0.09 & 1.52 $\pm$ 0.07 & -0.05 $\pm$ 0.14 \\
J004441.04-732136.4$^{c}$ & -1.34 $\pm$ 0.17 & 1.9 $\pm$ 0.7 & 2.06 $\pm$ 0.10 & 2.58 $\pm$ 0.10 & 2.40 $\pm$ 0.10 & 0.52 $\pm$ 0.10 \\   
J050632.10-714229.9$^{d}$ & -1.15 $\pm$ 0.17 & 1.5 $\pm$ 0.3 & 1.42 $\pm$ 0.07 & 1.07 $\pm$ 0.14 & 1.19 $\pm$ 0.09 & -0.35 $\pm$ 0.15 \\  
J051213.81-693537.1$^{e}$ & -0.56 $\pm$ 0.15 & 1.3 $\pm$ 0.4 & 1.33 $\pm$ 0.08 & 1.74 $\pm$ 0.08 & 1.61 $\pm$ 0.06 & 0.41 $\pm$ 0.12 \\
J051848.86-700246.9$^{e}$ & -1.03 $\pm$ 0.14 & 1.3 $\pm$ 0.3 & 1.46 $\pm$ 0.10 & 2.12 $\pm$ 0.08 & 1.90 $\pm$ 0.07 & 0.66 $\pm$ 0.13 \\
J052043.86-692341.0$^{d}$ & -1.15 $\pm$ 0.17 & 1.6 $\pm$ 0.9 & 1.67 $\pm$ 0.12 & 1.85 $\pm$ 0.10 & 1.79 $\pm$ 0.08 & 0.19 $\pm$ 0.16 \\
J053250.69-713925.8$^{d}$ & -1.22 $\pm$ 0.11 & 2.5 $\pm$ 0.7 & 1.51 $\pm$ 0.16 & 1.94 $\pm$ 0.10 & 1.80 $\pm$ 0.08 & 0.43 $\pm$ 0.19 \\ 
\hline
\end{tabular}
    \begin{tablenotes}
      \small
      \item $^{(\textrm{a})}$ Galactic object, from \citet{vanwinckel00}. For C/O, we assume an uncertainty of 0.3 dex.
      \item $^{(\textrm{b})}$ Galactic object, from \citet{reyniers04}. For C/O, we assume an uncertainty of 0.3 dex.
      \item $^{(\textrm{c})}$ SMC object, from \citet{desmedt12}.
      \item $^{(\textrm{d})}$ LMC object, from \citet{vanaarle13}.
      \item $^{(\textrm{e})}$ LMC object, from this paper.      
    \end{tablenotes}
\end{threeparttable}
\end{table*}

\section{Neutron irradiation} \label{sect:neutron}

The \textit{s}-process distributions and \textit{s}-process overabundances are typically
represented by four observational indices: [ls/Fe], [hs/Fe], [s/Fe],
and [hs/ls]. Unfortunately, the elemental abundances used for the
calculations of these indices vary in the literature. To have a
significant statistical sample, we include the abundance results
from \citet{vanwinckel00} and \citet{reyniers04} for Galactic objects
in the \textit{s}-process index study. For the ls-index, we follow the
suggestion from \citet{busso95} and use the mean of the relative
abundances of Y and Zr. For the hs-index, normally the Ba abundance is
used, but since an accurate Ba abundance study is hampered by the
strong saturated Ba lines in \textit{s}-process enriched objects, we replace
the Ba abundance by the Ce abundance, which can be determined accurately in 
enriched objects. The hs-index is then the mean of the
relative abundances of La, Ce, Nd and Sm. Our four \textit{s}-process
indices are

\begin{equation*}
\textrm{[ls/Fe]} = \frac{\textrm{[Y/Fe]} + \textrm{[Zr/Fe]}}{2}; 
\end{equation*}
\begin{equation*}
\textrm{[hs/Fe]} = \frac{\textrm{[La/Fe]} + \textrm{[Ce/Fe]} + \textrm{[Nd/Fe]} + \textrm{[Sm/Fe]}}{4}; 
\end{equation*}
\begin{equation*}
\textrm{[s/Fe]} = \frac{\textrm{[(ls+hs)/Fe]}}{6};
\end{equation*}
\begin{equation*}
\textrm{[hs/ls]} = \textrm{[hs/Fe]} - \textrm{[ls/Fe]}.
\end{equation*}

Since we use a non-standard element selection for the hs index calculations, we
list the four \textit{s}-process indices of all stars in our \textit{s}-process indices
analysis in Table \ref{table:sindex}. This includes Galactic
object from \citet{vanwinckel00} and \citet{reyniers04}, and
Magellanic Cloud objects from this paper, \citet{desmedt12}, and
\citet{vanaarle13}. For objects IRAS 04296+3429, IRAS 19500-1709, and
J053250.69-713925.8, no Sm abundance could be determined. To estimate
the Sm abundance for these objects, we scale the abundance of Sm to
the abundance of Nd, which is the element with an atomic mass closest to
Sm and is included in the hs index. We use the AGB nucleosynthesis models
from the online-database FRUITY\footnote{http://fruity.oa-teramo.inaf.it/} 
\citep[Franec Repository
of Upgraded Isotopic Tables and Yields][]{cristallo11} and use
the models with metallicities closest to the respective star's
metallicity. We choose a standard mass of 1.5 M$_\odot$ based upon our
estimated initial mass results for the Magellanic Cloud objects (see Sect. \ref{subsect:inimass})

\begin{figure}[t!]
\resizebox{\hsize}{!}{\includegraphics{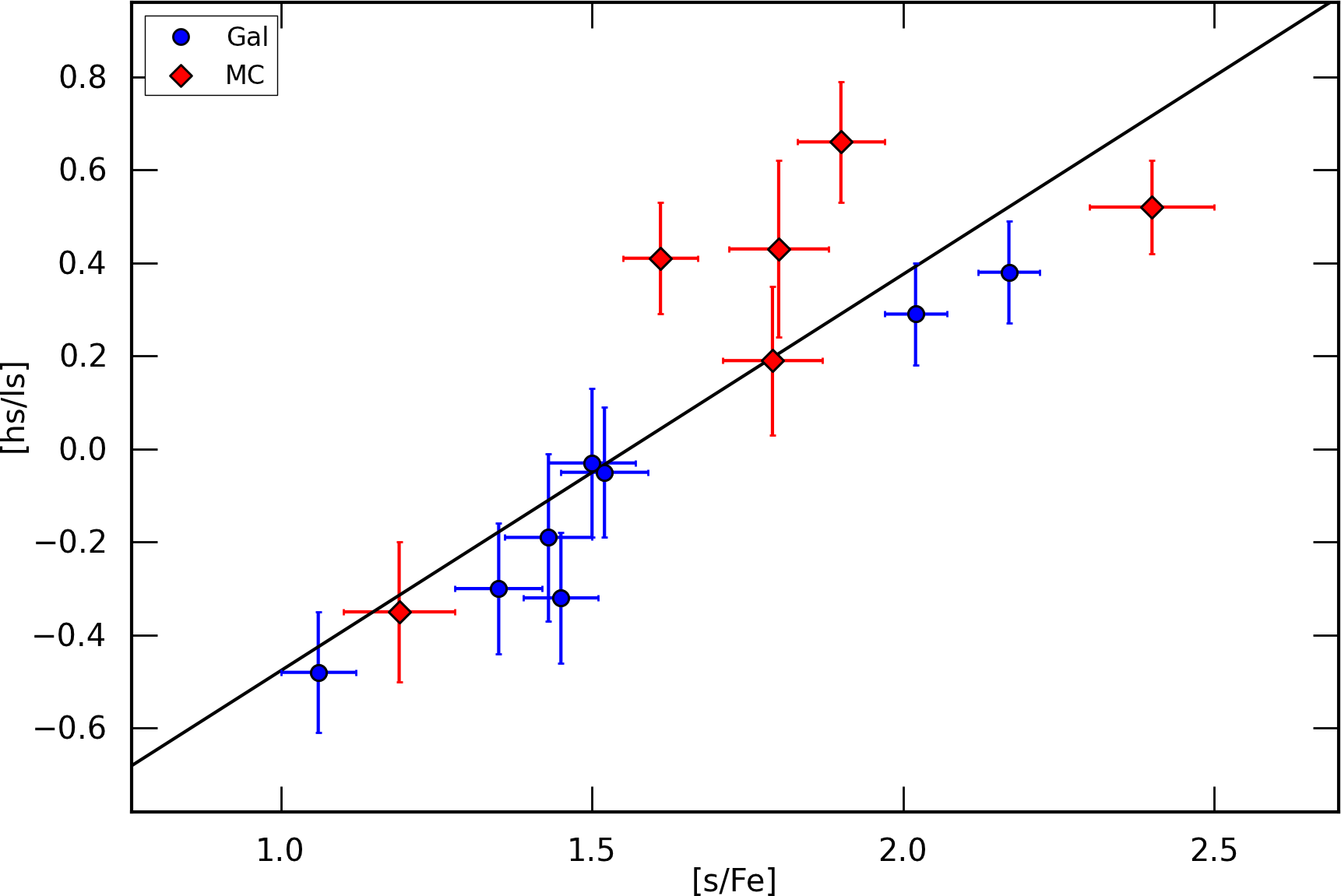}}
\caption{Correlation between the total enrichment in \textit{s}-process elements and the [hs/ls] index for the stars listed in \ref{table:sindex}. 
Galactic objects are represented by blue dots; Magellanic Cloud objects are represented by red diamonds. The full line shows the least-squares 
fit to all data points.}\label{fig:sfe}
\end{figure}

\begin{figure}[t!]
\resizebox{\hsize}{!}{\includegraphics{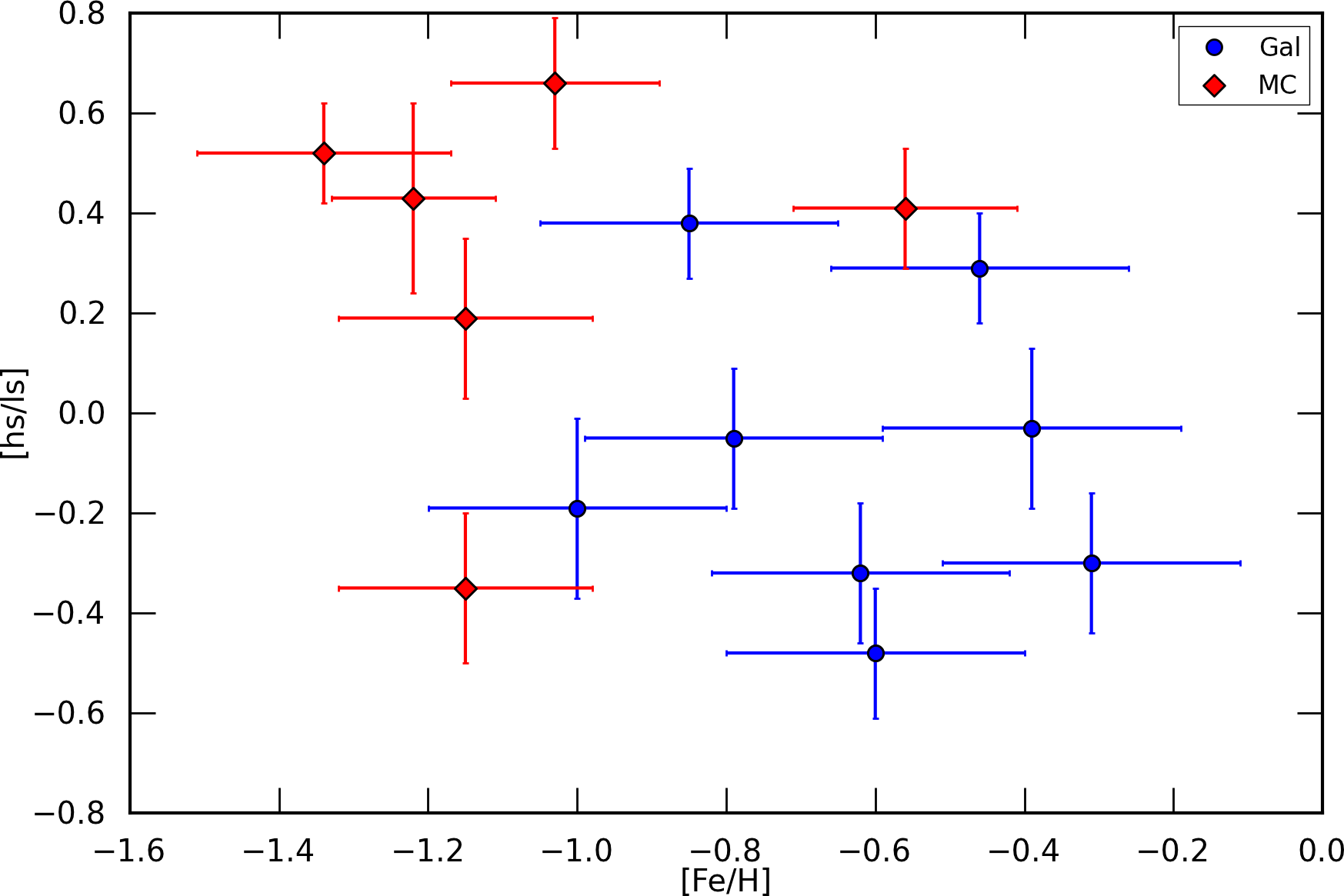}}
\caption{The absence of correlation between the metallicity [Fe/H] and the [hs/ls] index for all stars listed in \ref{table:sindex}. 
Symbols and the line are similar to Fig. \ref{fig:sfe}. There is no clear correlation between metallicity and this range of metallicities.}\label{fig:feh}
\end{figure}

\citet{vanwinckel00}, \citet{reyniers04}, and \citet{vanaarle13} find a
strong correlation between [s/Fe] and [hs/ls] indices in Galactic and
Magellanic Cloud objects, for which elements with high [s/Fe]
generally show a higher [hs/ls] ratio. We add three objects \citep[our two
sample stars and J004441 from ][]{desmedt12} to this relation (see
Fig. \ref{fig:sfe}). We find that these three objects confirm the
correlation between [s/Fe] and [hs/ls] with a correlation coefficient
of 0.86. The [s/Fe] index represents the third dredge-up efficiency,
although it is also influenced by the mass-loss history during the AGB
phase and the envelope mass during TDUs. Note that the [hs/ls] index represents
neutron irradiation, the number of neutrons that are available for each iron seed nucleus. 
Generally, an overabundance of hs-elements with
respect to ls-elements is expected in low-mass and low-metallicity AGB stars (1-3
M$_{\odot}$) in which the $^{13}$C($\alpha$,n)$^{16}$O is expected to
be the main neutron source for the \textit{s}-process
\citep[e.g.][]{straniero95, gallino98,abia02,cristallo11,karakas14}. In heavier
AGB stars, the $^{22}$Ne($\alpha$,n)$^{25}$Mg reaction is dominant,
which mainly produces ls-elements. Of all the Magellanic Cloud objects listed in
Table \ref{table:sindex}, all objects have low initial masses and an overabundance of hs
with respect to ls, except for J050632.10-714229.9 
(Fig. \ref{fig:inimass}).

As described above, the [hs/ls] index is expected to increase with
decreasing metallicity because of a larger number of neutrons for each
iron seed, assuming
that the diffusion of protons in the He-rich intershell is totally
independent of metallicity. In Fig. \ref{fig:feh}, 
which also includes our sample stars and J004441, there is no clear
correlation between [Fe/H] and [hs/ls]. 

\begin{figure}[t!]
\resizebox{\hsize}{!}{\includegraphics{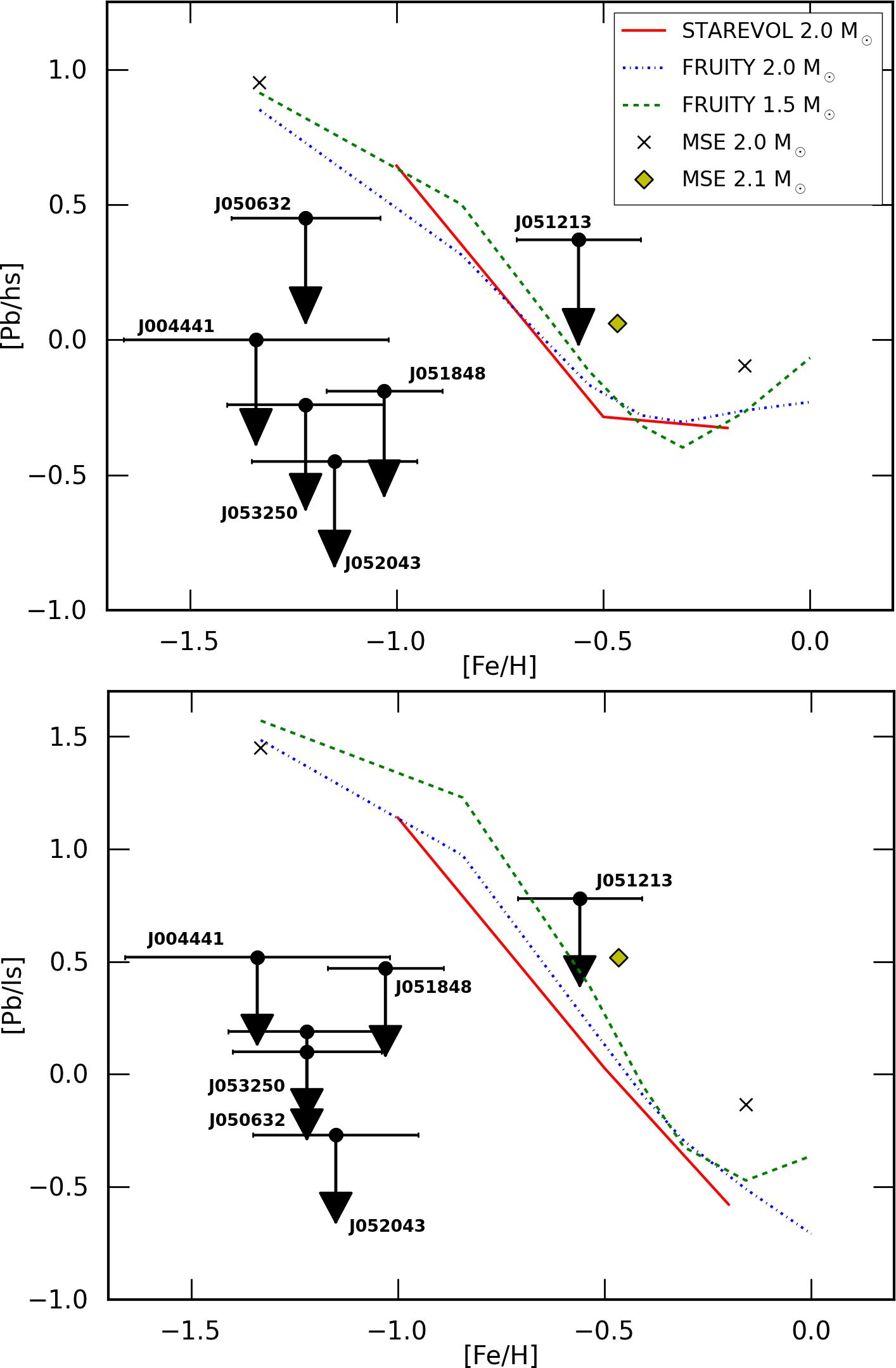}}
\caption{Overview of the observed [Pb/hs] (upper panel) and [Pb/ls] (lower panel) upper limits of 
\textit{s}-process enriched post-AGB stars in the Magellanic Clouds from this study and \citet{desmedt14}.
The observed abundance upper limits are plotted together with the [Pb/hs] and [Pb/ls] predictions 
of the 2.0 M$_{\odot}$ STAREVOL models (red full line),the 2.0 M$_{\odot}$ and 2.1 M$_{\odot}$
Mount Stromlo models (black crosses and yellow diamond, respectively) and the 1.5 and 2.0 M$_{\odot}$ 
FRUITY models (green dashed and blue dot-dashed lines, respectively). 
The black horizontal lines represent the [Fe/H] uncertainty of the  stars shown. 
The position of each star is indicated with the first part of its 2MASS name. 
For more information, see text.}\label{fig:pb_index}
\end{figure}

Low-mass, low-metallicity stars are expected to have high
overabundances of Pb with respect to the hs-elements. The AGB models
predict the $^{13}$C neutron source to be of primary origin, which means
that it is independent of the initial metallicity. For similar models
with different metallicity, the neutron production is predicted to be
similar, but the number of available Fe seeds is different. Per Fe seed
nucleus, there are more neutrons available at low metallicity.
This results in the creation of heavier \textit{s}-elements. Since Pb
is the expected end-product of the \textit{s}-process nucleosynthesis,
high Pb abundances are expected in metal-poor stars.

In order to compare our Pb abundance results to mean theoretical predictions, we obtained 
predicted values from different AGB evolution and nucleosynthesis codes. For this comparison,
we use [Pb/hs] ([Pb/Fe] - [hs/Fe]) and [Pb/ls] ([Pb/Fe] - [ls/Fe]), which represent the overabundance 
of Pb with respect to the other \textit{s}-process elements. Fig. \ref{fig:pb_index} shows the comparison
between the observed [Pb/hs] (upper panel) and [Pb/ls] (lower) upper limits, together with 
2.0 M$_{\odot}$ model predictions of the STAREVOL code \citep[][and references therein]{siess07} 
for metallicities of [Fe/H] = -1.0, -0.5 and -0.2 dex (red full line), Mount-Stromlo Evolutionary code
(MSE) predictions \citep[][and references therein]{fishlock14,karakas10a} for [Fe/H] $\approx$ -1.3 and 
-0.15 dex for 2.0 M$_{\odot}$ models (black crosses) and [Fe/H] $\approx$ -0.5 dex for a 2.1 M$_{\odot}$ model 
(yellow diamond), and the publicly-available 2.0 M$_{\odot}$ model predictions of FRUITY \citep{cristallo11} within a 
metallicity range from [Fe/H] = -1.5 up to 0.0 dex (blue point-dashed line) for seven different metallicities. 
Because of the limited number of available metallicities within the shown metallicity ranges, the STAREVOL and FRUITY 
curves in Fig. \ref{fig:pb_index} are not smooth. We remark that, although all of the stars in Fig. \ref{fig:pb_index} 
have estimated initial masses between 1.0 and 1.5 M$_{\odot}$ 
\citep[see][and Sect. \ref{subsect:inimass}]{desmedt12,vanaarle13},
it is justified to make a comparison with 2.0 M$_{\odot}$ model predictions since the differences between the 1.5 
and 2.0 M$\sun$ FRUITY model predictions is only marginal. 

Fig. \ref{fig:pb_index} shows that the predictions fit well the
observed Pb abundance upper limit of J051213 but strongly overestimate
the Pb overabundances of the lower metallicity stars. There is 
an increasing discrepancy between predicted and observed Pb abundances
towards lower metallicities. A larger sample of post-AGB stars with
derived Pb abundances is required to confirm the observed discrepancy
towards lower metallicities. More research is clearly needed
to explain the consistent low Pb abundances in \textit{s}-process enriched low-metallicity 
post-AGB stars. This is beyond the scope of this study.

\section{Conclusions}\label{sect:conc}
We aim to understand the chemical diversity demonstrated by post-AGB stars,
and  we focus on \textit{s}-process rich objects. With detailed studies of 
a well-sampled grid of post-AGB objects covering a wide range in luminosity 
(hence core-mass and initial mass) and metallicity, we aim to build up a systematic set of 
constraints on AGB models. These will help us to understand the physical processes 
related to AGB nucleosynthesis and mixing. To obtain these constraints, 
we study newly discovered post-AGB stars in the Magellanic clouds. Because of 
their known distances, accurate luminosity and, hence, initial 
mass estimates can be made, which is contrary to the situation for Galactic peers.

Here, we report on a detailed chemical study of two
\textit{s}-process enriched optically visible post-AGB stars in the LMC. The objects
were carefully selected from our low-resolution surveys
\citep[][accepted]{vanaarle11,kamath15_accepted}. Our high-resolution spectral data reveal
very rich spectra, literally swamped with atomic lines of \textit{s}-process
elements.  We quantified that both objects belong to the most
\textit{s}-process rich objects known to date despite their similarly low
C/O ratios. The metallicity difference between both stars is about 0.5 dex.

With the addition of our two sample stars, all
\textit{s}-process enhanced post-AGB stars in the LMC and SMC studied
thus far cluster in the same region of the HR-diagram. All are
associated with the latest evolutionary phase of low-mass stars with, on average, low metallicity.

We also confirm the correlation between [hs/ls] and [s/Fe] for
\textit{s}-process enriched post-AGB stars and find that both Galactic and
Magellanic Cloud objects follow the same relation. The neutron exposure as
traced by [hs/ls] is correlated with the overall \textit{s}-process
overabundances, as traced by [s/Fe]. [s/Fe] is also determined by the amount of dredged-up intershell 
material and the envelope dilution. In addition, we
strengthen and confirm that there is no clear correlation between neutron
irradiation and metallicity. Furthermore, we find an increasing discrepancy between observed 
and predicted Pb overabundances towards lower metallicities: the higher metallicity J051213 
with [Fe/H] $\approx$ -0.56 dex fits well the model predictions, while J051848 and other 
Magellanic Cloud objects from previous studies reveal much lower [Pb/hs] and [Pb/ls] results 
than predicted.

Our current and future research involves expanding our survey of chemical studies in the
Magellanic clouds to construct a grid that cover a larger and well-sampled range in
luminosity (hence initial mass) 
and metallicity. We intend to use this grid of constraints to fine-tune
AGB models and therefore the underlying physics of nucleosynthesis and mixing.

\begin{acknowledgements}
The authors thank the referee Claudio B. Pereira for the constructive comments that improved the quality of this paper. The authors thank L. Siess, S. Goriely
for providing the STAREVOL Pb abundance predictions. The authors thank A. I. 
Karakas for providing the Mount Stromlo Pb abundance predictions. KDS, HVW, and DK 
acknowledge support of the KU Leuven contract GOA/13/012. DK acknowledges support of 
the FWO grant G.OB86.13. 

\end{acknowledgements}

\bibliographystyle{aa}
\bibliography{allreferences}

\end{document}